\documentclass[%
 reprint,
 amsmath,amssymb,
 aps,
]{revtex4-2}

{}
{}
{}
\usepackage{url}
\usepackage{enumerate}
\usepackage{graphicx}
\usepackage{braket}
\usepackage{todonotes}
\usepackage{amsmath}
\usepackage{hyperref}
\usepackage{mathrsfs}
\usepackage{amsfonts}
\usepackage{mathtools}
\usepackage{comment}
\usepackage{longtable}
\DeclarePairedDelimiter\ceil{\lceil}{\rceil}
\usepackage{rotating}

\usepackage{graphicx}
\usepackage{dcolumn}
\usepackage{bm}

\usepackage[qm]{qcircuit}
\usepackage{amsmath}
\usepackage{tikz}

\newcommand*\qcontrolcolor[2]{\push{\footnotesize \tikz[baseline=(char.base)]{
                              \node[shape=circle,draw,inner sep=0.6pt,color=#1] (char) {#2};}}
                              \qw}  

\newcommand*\onecontrol{\qcontrolcolor{red!80!black}{1}}
\newcommand*\twocontrol{\qcontrolcolor{blue!100!white}{2}}
\newcommand*\threecontrol{\qcontrolcolor{black!100!white}{3}}
\newcommand*\dminusonecontrol{\qcontrolcolor{red!80!black}{$d$-1}}
\newcommand*\dcontrol{\qcontrolcolor{blue!100!white}{$d$}}
\newcommand*\dplusonecontrol{\qcontrolcolor{black!100!white}{$d$+1}}

\usepackage{tikz}

\usetikzlibrary{automata, positioning, arrows, decorations.markings}
\tikzset{myptr/.style={decoration={markings,mark=at position 1 with %
    {\arrow[scale=2,>=stealth]{>}}},postaction={decorate}}}

\definecolor{brown}{RGB}{77, 51, 0}

\begin{document}

\preprint{APS/123-QED}

\title{Asymptotically Improved Circuit for $d$-ary Grover's Algorithm with Advanced Decomposition of $n$-qudit Toffoli Gate}

\author{Amit Saha$^{1,2}$} 
\email{abamitsaha@gmail.com}
\author{Ritajit Majumdar$^3$}
\author{Debasri Saha$^{1}$}
\author{Amlan Chakrabarti$^{1}$}
\author{Susmita Sur-Kolay$^3$}

\affiliation{$^{1}$A. K. Choudhury School of Information Technology, University of Calcutta, India\\
$^{2}$ATOS, Pune, India\\$^3$Advanced Computing \& Microelectronics Unit, Indian Statistical Institute, India}

\date{\today}

\begin{abstract}
 The progress in building quantum computers to execute quantum algorithms has recently been remarkable. Grover's search algorithm in a binary quantum system provides considerable speed-up over classical paradigm. Further, Grover's algorithm can be extended to a $d$-ary (qudit) quantum system for utilizing the advantage of larger state space, which helps to reduce the run-time of the algorithm as compared to the traditional binary quantum systems. In a qudit quantum system, an $n$-qudit Toffoli gate plays a significant role in the accurate implementation of Grover's algorithm. In this article, a generalized $n$-qudit Toffoli gate has been realized using higher dimensional  qudits to attain a logarithmic depth decomposition without ancilla qudit. The circuit for Grover's algorithm has then been designed for any $d$-ary quantum system, where $d \ge 2$, with the  proposed $n$-qudit Toffoli gate to obtain optimized depth compared to earlier approaches. The technique for decomposing an $n$-qudit Toffoli gate requires access to two immediately higher energy levels, making the design susceptible to errors. Nevertheless, we show that the percentage decrease in the probability of error is significant as we have reduced both gate count and circuit depth as compared to that in state-of-the-art works.  
\end{abstract}

\maketitle


\section{Introduction}\label{sec1}

The proliferation of quantum algorithms is gradually grabbing the eye of researchers. Quantum computer hardware is now available for physical implementation of these algorithms to attain significant speedups \cite{chuang}. Conventionally, classical computers are designed on transistors, which deal with binary bits at the physical level. Quantum computers are designed to deal with qubit technology. Albeit, the fundamental physics behind the quantum systems is not inherently binary, on the contrary, a quantum system can have an infinite arity of discrete energy levels. In reality, the limitation lies in the fact that we need to control the system as per our needs. Including additional discrete energy levels for the purpose of computation helps us to realize the qudit technology quite comprehensively, which makes the system more flexible with data storage and faster in processing of quantum information.

 Qudit technology generally deals with $d$-ary quantum systems, where $d > 2$ \cite{brylinski}. For providing a larger state space and simultaneous multiple control operations, we consider qudits which eventually reduce the circuit complexity and enhance the efficiency of quantum algorithms \cite{qpe, qft, LI20114249, 9410395}. For example, $N$ qubits can be expressed as $\frac{N}{log_2 {d}}$ qudits, which shaves off by a $log_2 {d}$-factor from the run-time of a quantum algorithm \cite{universal, geometry}. The $d$-ary quantum computing system can be realized on various physical technologies, for instance, continuous spin systems \cite{Bartlett_2002,Adcock_2016}, superconducting transmon technology \cite{PhysRevA.76.042319}, nuclear magnetic resonance \cite{Dogra_2014, Gedik_2015}, photonic systems \cite{Gao_2020},
 ion trap \cite{qutrit}, topological quantum systems \cite{Cui_2015first, Cui_2015, bocharov2015improved} 
and molecular magnets \cite{Leuenberger_2001}. In this work, we consider the implementation of Grover's search algorithm \cite{Grover} generalised to a $d$-ary quantum system. The goal of $d$-ary Grover's search algorithm is to search data from an unstructured database, and attain significant speed-up compared to its classical counterpart. 

In this article, we have designed an efficient quantum circuit for Grover's algorithm using the proposed novel decomposition of an $n$-qudit Toffoli gate \cite{Muthukrishnan_2000}. For physical realization of $n$-qudit Toffoli gate, it is of utmost importance to decompose it into one-qudit and/or two-qudit gates. In \cite{gokhalefirst}, authors have proposed a qubit-qutrit approach to decompose a generalized Toffoli gate, which we have extended to $n$-qudit Toffoli decomposition with the use of $\ket{d}$ and $\ket{d+1}$ quantum states as temporary storage. We propose here an approach similar to that in  \cite{gokhalefirst} for extending the decomposition of generalized $n$-qudit Toffoli gate in terms of $d+1$-ary Toffoli gate. However, instead of decomposing $d+1$-ary Toffoli gate for simulation purpose, the $d+1$-ary Toffoli gate has been decomposed into $d+1$-ary and $d+2$-ary CNOT gates to achieve optimized depth. By simply adding a discrete energy level, we can easily have a higher dimension quantum state for temporary use, since these are present only as intermediate states in a qudit system, whereas the input and output states are qudits. In the intermediate operations alone, we introduce the $\ket{d}$ and $\ket{d+1}$ quantum state of $d+2$-ary quantum systems without hampering the operation of initialization and measurement on physical devices. By introducing the $d+2$-ary quantum systems, the constraint of multi-valued multi-controlled Toffoli decomposition can be avoided. To the best of our knowledge, it is a first of its kind approach. As the $d$-ary system may need to occasionally access states beyond the $d$-ary computational space ---  an engineering challenge, it makes the system particularly susceptible to error \cite{gokhale2019asymptotic}. We have shown the effect of generic noise models on the proposed implementation of Toffoli decomposition. 
 
  Our contributions are the following:
\begin{itemize}
    \item a novel technique to decompose a generalized $n$-qudit Toffoli gate into a $\log_2 n$ depth and no-ancilla qudit equivalent circuit --- as  an example, a 8-qubit Toffoli ($C^7NOT$) gate realization has been demonstrated and a comparative study depicts that our approach is better than the existing approaches in terms of a constant factor of gate cost reduction;
    \item a circuit for Grover's search algorithm achieves a logarithmic depth in any $d$-ary quantum system using the proposed decomposed $n$-qudit Toffoli gate as compared to linear depth;
    
    \item study of the effect of the generic error models (gate error and idle error) for the proposed decomposition, keeping aside the noise mitigation techniques which are not addressed here.
\end{itemize}

The layout of this article is as follows. Section 2 describes the universal qudit gates. Section 3 presents the $d$-ary Grover's search algorithm. Section 4 illustrates the decomposition of the proposed $n$-qudit Toffoli gate and its comparative analysis. The performance of the decomposition under various types of noise is presented in Section 5. Section 6 captures our conclusions.

\section{Generalized Qudit Gates}\label{sec2}

A \textit{qudit} is the unit of quantum information for $d$-ary quantum systems\cite{lanyon, Wang_2020}. Qudit states can be expressed by a
vector in the~$d$ dimensional Hilbert space~$\mathscr{H}_d$~\cite{universal, Kiktenko_2020}.
The vector space is the span of orthonormal basis vectors $\{\ket0,\ket1,\ket2,\dots \ket{d-1}\}$. 
The general form of qudit state can be described as 
\begin{equation}
\ket{\psi}=\alpha_0 \ket0 +\alpha_1 \ket1 +\alpha_2 \ket2+\cdots+\alpha_{d-1} \ket{d-1}=
\begin{pmatrix}
\alpha_0 \\
\alpha_1 \\
\alpha_2 \\
\vdots   \\
\alpha_{d-1} \\
\end{pmatrix}
\end{equation}
where $|\alpha_0|^2+|\alpha_1|^2+|\alpha_2|^2+\cdots+|\alpha_{d-1}|^2=1$ and $\alpha_0$, $\alpha_1$, $\dots$, $\alpha_{d-1} \in\mathbb{C}^d$. An overview of generalized qudit gates is presented in this section. The generalisation can be defined as discrete quantum states of any arity \cite{Bullock_2005}. Unitary qudit gates \cite{Daboul_2003, Jafarzadeh_2020} are applied on qudits  to  modify  the  quantum  state  in a quantum algorithm \cite{Barenco}. For logic  synthesis of  Grover's algorithm in $d$-ary quantum systems, one needs to consider  one-qudit generalized gates such as NOT gate ($X_d$),  Hadamard  gate ($F_d$), two-qudit generalized CNOT gate ($C_{X,d}$) and Generalized $n$-qudit Toffoli gate ($C^{n}_{X,d}$). These  gates are defined next.

\subsection{Generalized NOT Gate}

$X_d$ is the generalized NOT or increment gate  \cite{patera}.

\subsection{Generalized Hadamard Gate}

$F_d$ is the generalized quantum Fourier transform or generalized Hadamard gate  \cite{LI20114249, Fan_2007}, which produces the superposition of the input basis states.

\subsection{Generalized CNOT Gate}

Quantum entanglement is a phenomenal property of quantum mechanics, and can be achieved by a controlled NOT (CNOT) gate in binary quantum systems. For $d$-ary quantum systems, the binary two-qubit CNOT gate is generalised to $\text{$C_{X,d}$}\ket{x}\ket{y}=\ket{x}\ket{(y+1)\mod d}$ only if $x=d-1$, and otherwise = $\ket{x}\ket{y}$ \cite{Di_2013}.

\subsection{Generalized $n$-qudit Toffoli Gate}

Next, we extend the generalized CNOT further to operate over $n$ qudits as a generalized $n$-qudit Toffoli gate $C_{X,d}^n$ \cite{Muthukrishnan_2000}.  For $C_{X,d}^n$, the target qudit is incremented by $1 \ (\text{mod } d)$ only when all the $n-1$ control qudits are $d-1$. The $(d^n \times d^n)$ matrix representation of generalized $n$-qudit Toffoli gate is as follows:

\begin{equation*}
C_{X,d}^n = \left( \begin{matrix}
    I_d & 0_d & 0_d & \ldots & 0_d \\
    0_d & I_d & 0_d & \ldots & 0_d \\
    0_d & 0_d & I_d & \ldots & 0_d \\
    \vdots & \vdots & \vdots & \ddots & \vdots \\
    0_d & 0_d & 0_d & \ldots &  X_d \\
\end{matrix} \right)
  \end{equation*}
  
  where $I_d$ and $0_d$ are both $d \times d$ matrices as shown below:
\begin{equation*}
I_d = 
\begin{pmatrix}
    1 & 0 & 0 & \ldots & 0 \\
    0 & 1 & 0 & \ldots & 0 \\
    0 & 0 & 1 & \ldots & 0 \\
    \vdots & \vdots & \vdots & \ddots & \vdots \\
    0 & 0 & 0 & \ldots &  1 \\
\end{pmatrix} 
\quad\textrm{and}\quad
0_d =  
\begin{pmatrix}
    0 & 0 & 0 & \ldots & 0 \\
    0 & 0 & 0 & \ldots & 0 \\
    0 & 0 & 0 & \ldots & 0 \\
    \vdots & \vdots & \vdots & \ddots & \vdots \\
    0 & 0 & 0 & \ldots &  0 \\
\end{pmatrix}
\end{equation*} 

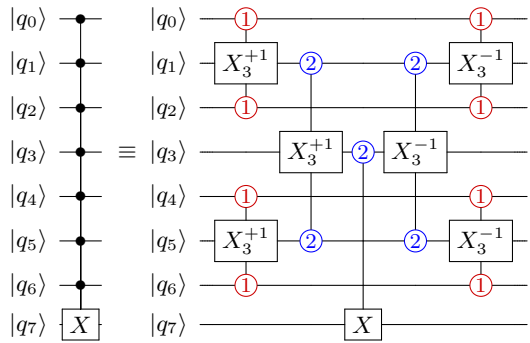
\begin{figure}[!htb]
\[
\Qcircuit @R=0.5em @C=0.1em {
\lstick{\ket{q_0}} & \ctrl{7} & \qw & &  & &  & &  & &  & &  & & & &  & &  & &  & &  & \lstick{\ket{q_{0}}} & \qw & \qw & \qw & \qw & \qw & \onecontrol & \qw  & \qw  & \qw    & \onecontrol & \qw & \qw & \qw & \qw & \qw & \qw \\
\lstick{\ket{q_1}} & \ctrl{6} & \qw & &  & &  & &  & &  & &  & &  & &  & & &  & &  & &  \lstick{\ket{q_{1}}} & \qw & \qw & \qw & \qw & \qw & \gate{X_3^{+1}}\qwx  & \twocontrol & \qw   & \twocontrol & \gate{X_3^{-1}}\qwx  & \qw & \qw & \qw & \qw & \qw & \qw \\
\lstick{\ket{q_2}} & \ctrl{5} & \qw & &  & &  & &  & &  & &  & & & &  & &  & &  & &  & \lstick{\ket{q_{2}}} & \qw & \qw & \qw & \qw & \qw & \onecontrol\qwx  & \qw \qwx  & \qw  & \qw \qwx   & \onecontrol\qwx  & \qw & \qw & \qw & \qw & \qw & \qw \\
\lstick{\ket{q_3}} & \ctrl{4} & \qw & \push{\rule{.3em}{0em}\equiv\rule{.3em}{0em}} &  & &  & &  & &  & &  & & & &  & &  & &  & &  & \lstick{\ket{q_{3}}} & \qw & \qw & \qw & \qw & \qw & \qw  & \gate{X_3^{+1}}\qwx  & \twocontrol &  \gate{X_3^{-1}}\qwx  & \qw  & \qw & \qw & \qw & \qw & \qw & \qw \\
\lstick{\ket{q_4}} & \ctrl{3} & \qw & &  & &  & &  & &  & &  & & & &  & &  & &  & &  & \lstick{\ket{q_{4}}} & \qw & \qw & \qw & \qw & \qw & \onecontrol & \qw \qwx  & \qw \qwx   & \qw \qwx  & \onecontrol & \qw & \qw & \qw & \qw & \qw & \qw \\
\lstick{\ket{q_5}} & \ctrl{2} & \qw & &  & &  & &  & &  & & & &  & &  & &  & &  & &  & \lstick{\ket{q_{5}}} & \qw & \qw & \qw & \qw & \qw & \gate{X_3^{+1}}\qwx  & \twocontrol\qwx  & \qw \qwx    & \twocontrol\qwx  & \gate{X_3^{-1}}\qwx  & \qw & \qw & \qw & \qw & \qw & \qw \\
\lstick{\ket{q_6}} & \ctrl{1} & \qw & &  & &  & &  & &  & &  & & & &  & &  & &  & &  & \lstick{\ket{q_{6}}} & \qw & \qw & \qw & \qw & \qw & \onecontrol\qwx  & \qw  & \qw \qwx   & \qw  & \onecontrol\qwx  & \qw & \qw & \qw & \qw & \qw & \qw \\
\lstick{\ket{q_7}} & \gate{X} & \qw &
 & & &  & &  & &  & & & &  & &  & &  & &  & &  & \lstick{\ket{q_{7}}} & \qw & \qw & \qw & \qw & \qw & \qw  & \qw  & \gate{X}\qwx   & \qw  & \qw  & \qw & \qw & \qw & \qw & \qw & \qw \\
}
\]

\caption{Decomposition of an 8-qubit Toffoli gate with an intermediate qutrit \cite{gokhale2019asymptotic}. Each input and output is a qubit. The red control qubits activate on $\ket{1}$
and the blue controls activate on $\ket{2}$. In the rectangle for the target qubit $X^{+1}_3$ and $X^{-1}_3$ denote the $modulo~3$ increment and decrement operation respectively.}
\label{gokhale8}

\end{figure}

Due to technology constraints, a multi-controlled Toffoli gate can be replaced by an equivalent circuit comprising one-qudit and/or two-qudit gates, albeit at first the multi-controlled Toffoli has to be decomposed into a set of Toffoli gates for an arbitrary finite-dimensional quantum system \cite{Khan_2006}. In FIG. \ref{gokhale8}, we have shown an example of the state-of-the-art approach of decomposition of an 8-qubit Toffoli gate with the help of an intermediate qutrit state \cite{gokhale2019asymptotic, 10.1145/3406309}. The equivalent circuit temporarily stores information directly in the qutrit  state $\ket{2}$ as the controls,which are marked in blue color.

\begin{figure*}

\[
\Qcircuit @R=0.5em @C=0.25em {
\lstick{\ket{q_{0}}} & \qw  & \qw  & \qw  & \onecontrol & \qw  & \qw  & \qw  & \qw  & \qw  & \onecontrol & \qw  & \qw  & \qw  & \rstick{\ket{q_{0}}} \qw & \\
\lstick{\ket{q_{1}}} & \qw  & \onecontrol & \qw  & \qw \qwx  & \qw  & \onecontrol & \qw  & \onecontrol & \qw  & \qw \qwx  & \qw  & \onecontrol & \qw  & \rstick{\ket{q_{1}}} \qw & \\
\lstick{\ket{q_2}} & \gate{R_{y, 01}(\frac{-\pi}{4})} & \gate{X_3^{12}}\qwx  & \gate{R_{y, 01}(\frac{-\pi}{4})} & \gate{X_3^{12}}\qwx  & \gate{R_{y, 01}(\frac{\pi}{4})} & \gate{X_3^{12}}\qwx  & \gate{R_{y, 01}(\frac{\pi}{4}) \cdot R_{y, 12}(\frac{\pi}{4})} & \gate{X_3^{01}}\qwx  & \gate{R_{y, 01}(\frac{\pi}{4})} & \gate{X_3^{01}}\qwx  & \gate{R_{y, 01}(\frac{-\pi}{4})} & \gate{X_3^{01}}\qwx  & \gate{R_{y, 01}(\frac{-\pi}{4})} & \rstick{\ket{q_0 \wedge q_1}} \qw & \\
}
\]
\caption{Decomposition of ternary Toffoli gate into 13 one-qutrit and two-qutrit gates \cite{Di_2013}.}
\label{toffoli_ternary}

\end{figure*}
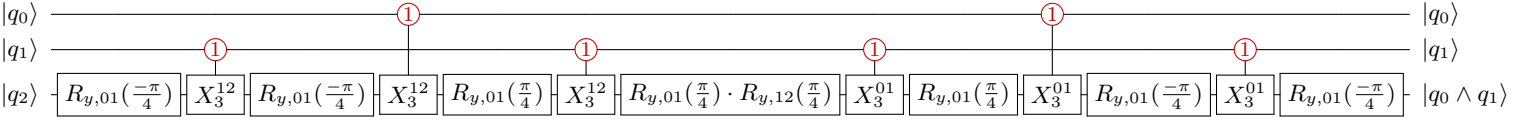

In the schematic diagram of the circuit, a circle denotes a  control qubit, and a rectangle the target qubit. As shown in FIG. \ref{gokhale8}, each of the two circles for the two control qubits of the binary Toffoli gates in the first level has $1$ marked in it with blue color, and `$X^{+1}_3$' in the rectangle for the target qubit to represent the $modulo~3$ increment operation. 
The decomposed circuit can be treated as a binary tree of gates which establishes the logarithmic depth for a multi-controlled Toffoli gate. It has the property that the intermediate qubit of each sub-tree as well as the root can only be raised to $\ket{2}$ if all of its seven control leaves are $\ket{1}$. In order to verify this property, we perceive that the qubit $q_3$ can only become $\ket{2}$ if and only if it was originally $\ket{1}$, and $q_1$ and $q_5$ qubits were previously $\ket{2}$. Then at the subsequent level of the tree, we observe that (i) qubit $q_1$ could have  been $\ket{2}$ only if it was previously $\ket{1}$, and both $q_0$ and $q_2$ were $\ket{1}$ earlier, (ii) qubit $q_5$ could have  been $\ket{2}$ only if it was previously $\ket{1}$ and both $q_4$ and $q_6$ qubits were $\ket{1}$ earlier. If any of the controls were not $\ket{1}$, the $\ket{2}$ state would fail to move to the root of the tree. Hence, the $CNOT$ gate toggles the target qubit  only  if all controls are $\ket{1}$. The right half of the circuit is the mirror circuit to restore the control qubits to their original states. The authors in  \cite{Di_2013} have further decomposed their ternary Toffoli gate into 13 one-qutrit and two-qutrit gates for physical implementation as shown in FIG. \ref{toffoli_ternary}.

These works are nonetheless restricted to binary quantum systems as they have mentioned the use of qutrits only. In our proposed approach, we have generalized the decomposition for an arbitrary finite-dimensional quantum system.  

\section{Generalized Grover's Algorithm in $d$-ary quantum systems}\label{sec3}

We present the details of our proposed generalized Grover's algorithm in $d$-ary quantum systems here. The algorithm has two sub-parts: Oracle and diffusion \cite{Grover}. Formally, Grover's algorithm for searching in an unstructured database can be defined as follows: given a collection of unstructured database elements $x=1, 2 ,\ldots, N$, and an \emph{Oracle function} $f(x)$ that acts on a marked element $s$ as follows \cite{groverd},
\begin{equation}
f(x) =\left\{\begin{array}{ll}
1,& x=s, \\
0,& x\neq s,
\end{array}\right.
\label{oracle_fun}
\end{equation}
find the marked element with as few calls to $f(x)$ as possible \cite{Grover, groverd}. The database is encoded into a superposition of quantum states with each element being assigned to a corresponding basis state. Grover's algorithm searches over each possible outcome, which is represented as a basis vector $\ket{x}$ in an $n$-ary Hilbert space in $d$-ary quantum systems. Correspondingly, the marked element is encoded as $\ket{s}$. Thus, after applying unitary operations as an oracle function to the \emph{superposition} of the different possible outcomes, the search can be done in parallel. Then the generalized diffusion operator, which is also known as inversion about the average operator, amplifies the amplitude of the marked state to increase its measurement probability using constructive interference, with simultaneous attenuation of all other amplitudes, and searches the marked element in $O(\sqrt{N})$ steps, where $N=d^n$ \cite{groverd}.

\begin{figure}[!h]

\[
\Qcircuit @R=0.75em @C=1em {
\lstick{} & \qw & \multigate{4}{\mbox{Oracle}} & \gate{F_d} & \gate{X_d} & \qw & \dminusonecontrol & \qw & \gate{X_d^{-1}} & \gate{F_d} & \rstick{} \qw & \\
\lstick{} & \qw & \ghost{\mbox{Oracle}} & \gate{F_d} & \gate{X_d} & \qw & \dminusonecontrol\qwx & \qw  & \gate{X_d^{-1}} & \gate{F_d} & \rstick{} \qw & \\
\lstick{} & \push{\substack{\begin{tikzpicture}\filldraw[black](0em,-0.8em)circle(0em);\filldraw[black](0em,-0.4em)circle(0.03em);\filldraw[black](0em,0em)circle(0.03em);\filldraw[black](0em,0.4em)circle(0.03em);\filldraw[black](0em,0.8em)circle(0em);\end{tikzpicture}}} & *+<1em,.9em>{\hphantom{\mbox{Oracle}}} & \push{\substack{\begin{tikzpicture}\filldraw[black](0em,-0.8em)circle(0em);\filldraw[black](0em,-0.4em)circle(0.03em);\filldraw[black](0em,0em)circle(0.03em);\filldraw[black](0em,0.4em)circle(0.03em);\filldraw[black](0em,0.8em)circle(0em);\end{tikzpicture}}} & \push{\substack{\begin{tikzpicture}\filldraw[black](0em,-0.8em)circle(0em);\filldraw[black](0em,-0.4em)circle(0.03em);\filldraw[black](0em,0em)circle(0.03em);\filldraw[black](0em,0.4em)circle(0.03em);\filldraw[black](0em,0.8em)circle(0em);\end{tikzpicture}}} &
\push{\substack{\begin{tikzpicture}\filldraw[black](0em,-0.8em)circle(0em);\filldraw[black](0em,-0.4em)circle(0.03em);\filldraw[black](0em,0em)circle(0.03em);\filldraw[black](0em,0.4em)circle(0.03em);\filldraw[black](0em,0.8em)circle(0em);\end{tikzpicture}}} & \push{\substack{\begin{tikzpicture}\filldraw[black](0em,-0.8em)circle(0em);\filldraw[black](0em,-0.4em)circle(0.03em);\filldraw[black](0em,0em)circle(0.03em);\filldraw[black](0em,0.4em)circle(0.03em);\filldraw[black](0em,0.8em)circle(0em);\end{tikzpicture}}}\qwx  &
\push{\substack{\begin{tikzpicture}\filldraw[black](0em,-0.8em)circle(0em);\filldraw[black](0em,-0.4em)circle(0.03em);\filldraw[black](0em,0em)circle(0.03em);\filldraw[black](0em,0.4em)circle(0.03em);\filldraw[black](0em,0.8em)circle(0em);\end{tikzpicture}}} & \push{\substack{\begin{tikzpicture}\filldraw[black](0em,-0.8em)circle(0em);\filldraw[black](0em,-0.4em)circle(0.03em);\filldraw[black](0em,0em)circle(0.03em);\filldraw[black](0em,0.4em)circle(0.03em);\filldraw[black](0em,0.8em)circle(0em);\end{tikzpicture}}} & \push{\substack{\begin{tikzpicture}\filldraw[black](0em,-0.8em)circle(0em);\filldraw[black](0em,-0.4em)circle(0.03em);\filldraw[black](0em,0em)circle(0.03em);\filldraw[black](0em,0.4em)circle(0.03em);\filldraw[black](0em,0.8em)circle(0em);\end{tikzpicture}}} & \rstick{} & \\
\lstick{} & \qw & \ghost{\mbox{Oracle}} & \gate{F_d} & \gate{X_d} & \qw & \dminusonecontrol\qwx & \qw  & \gate{X_d^{-1}} & \gate{F_d} & \rstick{} \qw & \\
\lstick{} & \qw & \ghost{\mbox{Oracle}} & \gate{F_d} & \gate{X_d} & \gate{F_d} & \gate{X_d}\qwx & \gate{F_d}  & \gate{X_d^{-1}} & \gate{F_d} & \rstick{} \qw & \\
}
\]

\caption{Generalized Circuit for Grover's Operator in $d$-ary quantum systems \cite{groverd}.}
\label{diff}
\end{figure}
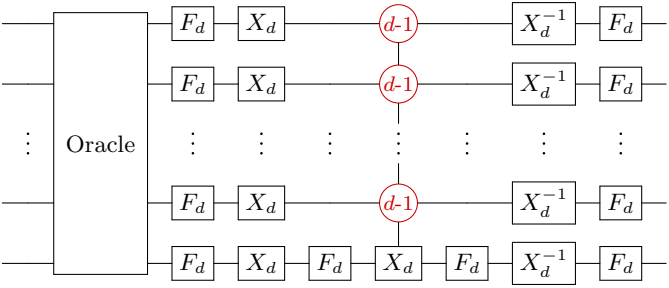

The generalized circuit for Grover's operator, the combination of the oracle and the diffusion in $d$-ary quantum systems is shown in FIG. \ref{diff}. As portrayed, the diffusion operator can be constructed using generalized Hadamard gate, generalized NOT gate and generalized $n$-qudit Toffoli gate. As discussed in Section 2, for implementing Grover's algorithm in technology specific physical devices, the $n$-qudit Toffoli gate needs to be decomposed using one-qudit or two-qudit gates. While decomposing the $n$-qudit Toffoli gate, if the depth and the ancilla qudits increase arbitrarily then the time complexity of Grover's algorithm also increases, which is undesirable. In the next subsection, we have shown a novel approach for the decomposition of an $n$-qudit Toffoli gate with optimized depth as compared to the state-of-the-art. Thus, the circuit depth of Grover's algorithm is also optimized.


\section{Improved Circuit for $d$-ary Grover's Search}\label{sec4}

In order to execute Grover's algorithm on quantum devices, it has to be ideally decomposed using one-qudit and/or two-qudit gates. It is important to carry out the effective low depth and low gate count decomposition in near term quantum devices and beyond \cite{Preskill_2018}. 

\subsection{Proposed Decomposition of $n$-qudit Toffoli Gate }\label{section3a}
 The most important aspect of our proposed work is the decomposition of $n$-qudit $d$-ary Toffoli gate. In the decomposition of generalized Toffoli gate, all the figures below have inputs and outputs as $d$-ary qudits, but the states $\ket{d}$ and $\ket{d+1}$ may be used in intermediate levels during the computation. The idea of keeping $d$-ary input/output enables these circuit constructions to be applied for any already existing $d$-ary qudit-only circuits.

\begin{figure}[!htb]
\centering
\[
    \Qcircuit @R=0.5em @C=0.25em {
\push{\rule{3em}{0em}} & \lstick{\ket{q_0}} & \ctrl{2} & \qw & &  & &  & &  & &  & &  & &   \lstick{\ket{q_0}} & \dminusonecontrol & \qw  & \dminusonecontrol  & \qw & \\
\push{\rule{3em}{0em}} & \lstick{\ket{q_1}} & \ctrl{1} & \qw &
\push{\rule{.3em}{0em}\equiv\rule{.3em}{0em}} & & &  & &  & &  & &  & &   \lstick{\ket{q_1}} & \gate{X^{+1}_{d+1}}\qwx  & \dcontrol  & \gate{X^{-1}_{d+1}}\qwx  &  \qw & \\
\push{\rule{3em}{0em}} &  \lstick{\ket{q_2}} & \gate{X_d} & \qw & &  & &  &   & &  & &  & &  & \lstick{\ket{q_2}} & \qw  & \gate{X_d} \qwx  & \qw  &  \qw & \\
    }
\]
\caption{Generalized Toffoli in $d$-ary quantum systems.}
\label{gentofdec}
\end{figure}
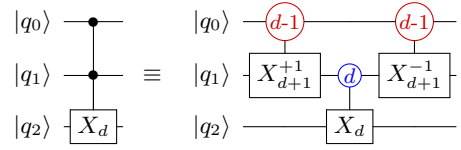

 A generalized Toffoli decomposition in a $d$-ary system using $\ket{d}$  state is shown in FIG. \ref{gentofdec}. A similar construction for the Toffoli gate in binary using qutrit is evident from previous state-of-the-art work \cite{gokhale2019asymptotic}; we have extended it for $d$-ary quantum systems. The aim is to carry out an $X_d$ operation on the target qudit (third qudit) as long as the two control qudits, are both $\ket{d-1}$. First, a $\ket{d-1}$-controlled $X^{+1}_{d+1}$, where $+1$ and $d+1$ are used to denote that the target qudit is incremented by $1 \ (\text{mod } d+1)$,  is performed on the first and the second qudits. This upgrades the second qudit to $\ket{d}$ if and only if the first and the second qudits were both $\ket{d-1}$. Then, a $\ket{d}$-controlled $X_d$ gate is applied to the target qudit. Therefore, $X_d$ is executed only when both the first and the second qudits were $\ket{d-1}$, as expected. The controls are reinstated to their original states by a $\ket{d-1}$-controlled $X^{-1}_{d+1}$ gate, which reverses the effect of the first gate. That the $\ket{d}$ state from $d+1$-ary quantum systems can be used instead of ancilla to store temporary information, which is the most important aspect in this decomposition.

As in \cite{gokhale2019asymptotic}, the circuit decomposition of generalized Toffoli gate is realized in terms of ternary Toffoli gate instead of one-qutrit and two-qutrit gates in order to obtain lower circuit depth. But during simulation, they decomposed the ternary Toffoli gate into six two-qutrit and seven one-qutrit physically implementable quantum gates. We have also followed a similar approach for extending the decomposition of generalized $n$-qudit Toffoli gate in terms of $d+1$-ary Toffoli gate. But, the approach of further  decomposition of the Toffoli for simulation purpose has not been adopted. Instead, the $d+1$-ary Toffoli gate has been decomposed into $d+1$-ary and/or $d+2$-ary CNOT gates. Let us consider a generalized CNOT gate for $d+2$-ary quantum systems as $C^{+1}_{X,d+2}$, where $+1$ and $d+2$ denote that the target qudit is incremented by $1 \ (\text{mod } d+2)$ only when the control qudit value is $d+1$. The $((d+2)^2 \times (d+2)^2)$ matrix representation of the $C^{+1}_{X,d+2}$ gate is as follows:

\begin{equation*}
C^{+1}_{X,d+2} = \left( \begin{matrix}
    I_{d+2} & 0_{d+2} & 0_{d+2} & \ldots & 0_{d+2} \\
    0_{d+2} & I_{d+2} & 0_{d+2} & \ldots & 0_{d+2} \\
    0_{d+2} & 0_{d+2} & I_{d+2} & \ldots & 0_{d+2} \\
    \vdots & \vdots & \vdots & \ddots & \vdots \\
    0_{d+2} & 0_{d+2} & 0_{d+2} & \ldots  &  X^{+1}_{d+2} \\
\end{matrix} \right)
  \end{equation*}

where $X^{+1}_{d+2}$ and $0_{d+2}$ are both $(d+2) \times (d+2)$ matrices as shown below:

\begin{align*}
X^{+1}_{d+2} = 
\left(\begin{matrix} 
0 & 0 & \ldots & 0 & 1 \\ 1 & 0 & \ldots & 0 & 0 \\ 0 & 1 & \ldots & 0 & 0 \\ \vdots & \vdots & \ddots & \vdots & \vdots \\ 0 & 0 & \ldots & 1 & 0
\end{matrix}\right)
\quad\textrm{and,}\quad
0_{d+2} = 
\left(\begin{matrix} 
0 & 0 & \ldots & 0 & 0 \\ 0 & 0 & \ldots & 0 & 0 \\ 0 & 0 & \ldots & 0 & 0 \\ \vdots & \vdots & \ddots & \vdots & \vdots \\ 0 & 0 & \ldots & 0 & 0
\end{matrix}\right)
  \end{align*}

\begin{figure*}[!htb]
\centering
\includegraphics[width=160mm,height=10cm]{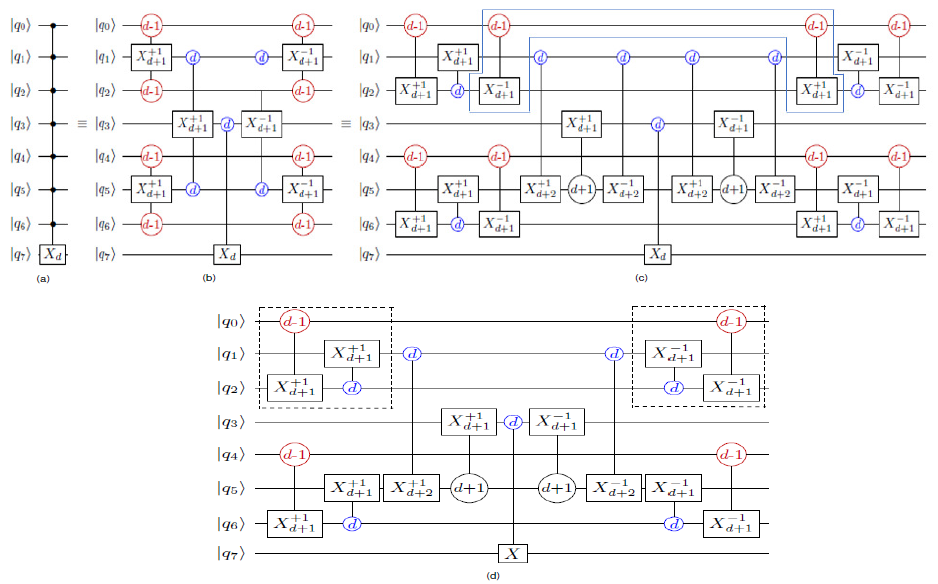}
\caption{ (a) An 8-qudit Toffoli gate, (b) its decomposition which follows \cite{gokhale2019asymptotic}, (c) our proposed decomposition using a few $d+1$-ary and/or $d+2$-ary CNOT gates, and (d) our proposed optimized decomposition.}
\label{8quditdecompose}
\end{figure*}

For example, let there be an 8-qudit Toffoli gate as shown in FIG. \ref{8quditdecompose}(a). First, we decompose it as in \cite{gokhale2019asymptotic} as shown in FIG. \ref{8quditdecompose}(b). Further, we decompose all the $d+1$-ary Toffoli gate into $d+1$-ary and/or $(d+2)$-ary CNOT gates as shown in FIG. \ref{8quditdecompose}(c) with the help of the proposed decomposition of generalized Toffoli in any arbitrary finite-dimensional quantum system. As shown in FIG. \ref{8quditdecompose}(c), all the $d-1$-controlled Toffoli gates are decomposed into $d-1$-controlled and $d$-controlled CNOT gates. Similarly, all the $d$-controlled Toffoli gates are decomposed into $d$-controlled and $d+1$-controlled CNOT gates. Thus, with the help of $\ket{d}$ and $\ket{d+1}$ quantum state of $(d+2)$-ary system, $X_d$ is executed if all the controlled qudits are in $\ket{d-1}$ state. In this manner, an $n$-qudit Toffoli gate can be decomposed. Further, the optimized $n$-qudit Toffoli gate decomposition has been portrayed in FIG. \ref{8quditdecompose}(d),  where two  generalized CNOT gates operating one after another on the same qudits are removed by applying the optimization rule as described in \cite{Nam_2018}. As an example, in FIG. \ref{8quditdecompose}(c), the two generalized CNOT gates marked within a {\it blue} boundary, have been eliminated by using identity rule, as no other gates are involved in between these two generalized CNOT gates on the $q_0$ and $q_2$. Hence, in FIG. \ref{8quditdecompose}(d), we conclude that for each generalized Toffoli decomposition, two generalized CNOT gates are sufficient for our proposed $n$-qudit Toffoli decomposition. For better understanding, as for example, we have highlighted the corresponding decomposition of FIG. \ref{8quditdecompose}(c), highlighted with {\it blue} boundary, in FIG. \ref{8quditdecompose}(d) with dotted boundary.  Now, if we want to apply our approach to a binary quantum system, then it could be easily carried out if a quaternary (4-ary quantum) \cite{sbm} system can also come into play. Moreover, we have achieved logarithmic depth as well as reduced the constant factor from $13$  to $2$, which is thoroughly discussed in the next subsection with the help of an example. By simulation, we have verified our circuits. The simulation results for the 8-qubit Toffoli gate of FIG. \ref{8qubit}, appears in the Appendix. In Table \ref{simu} of Appendix, we have shown the input and output states as well as intermediate states for each time cycle of the circuit for all possible combination of input states $\ket{00000000}$, $\ket{00000010}$, $\ket{00000100}$, \dots, $\ket{11111110}$. We have shown that only for the input state $\ket{11111110}$, the output state changes to $\ket{11111111}$, otherwise there is no change of output states for corresponding input states. 

\begin{figure*}[!htb]
\centering
\includegraphics[width=170mm,height=10cm]{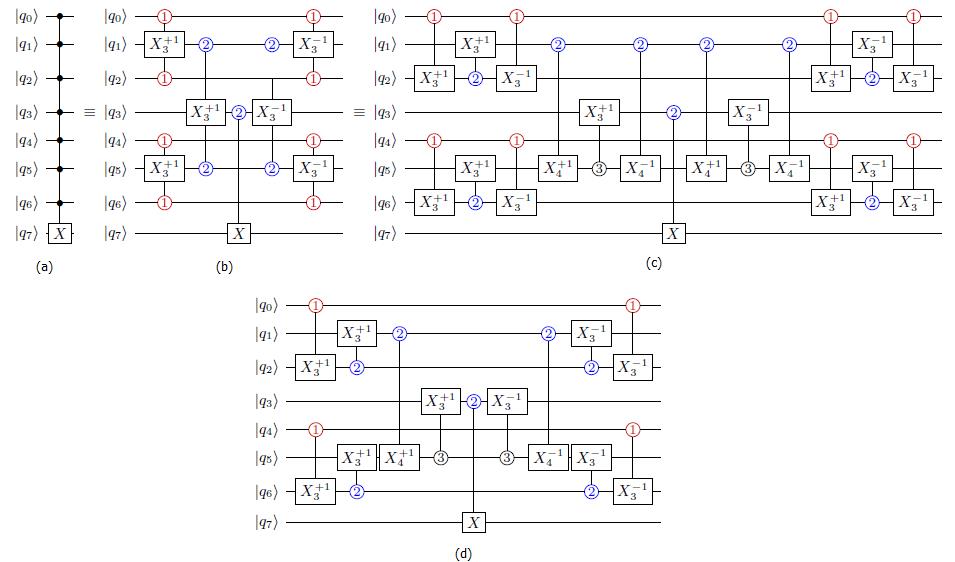}
\caption{(a) An 8-qubit Toffoli gate, (b) its decomposition in \cite{gokhale2019asymptotic}, (c) our proposed decomposition using a few ternary and/or quaternary CNOT gates, and (d) our proposed optimized decomposition..}
\label{8qubit}
\end{figure*}

\subsection{Comparative Analysis of Cost of Decomposition}\label{section3b}

A comparative study of our multi-controlled Toffoli decomposition with some previous works \cite{gokhale2019asymptotic, he, lanyon, wang, gidney, Barenco, Ralph_2007} is shown in Table \ref{tab:n_controlled}. Our work outperforms all of them in terms of depth of the circuit, even the best till now \cite{gokhale2019asymptotic}. We simulate our work taking the conventional construction proposed by Gokhale et al. \cite{gokhalefirst} into account, since it is the benchmark in the ancilla-free frontier zone. The technique makes the decomposition typically exorbitant in gate count and depth as a large number for constant factor of gate-count is required as compared to our approach. This is better explained with the following example.

\begin{table*}[!htb]
\caption{Asymptotic comparison of $n$-controlled Toffoli gate decomposition}
\centering
\resizebox{16cm}{!}{%
\begin{tabular}{|c|c|c|c|c|c|c|c|}
\hline
  & \textbf{This Work} & Gokhale\cite{gokhale2019asymptotic} & Gidney \cite{gidney} & He \cite{he} & Barenco \cite{Barenco} & Wang \cite{wang} & Lanyon \cite{lanyon}, Ralph \cite{Ralph_2007} \\ \hline
 Depth & $\log_2{n}$  & $\log_2{n}$ & $n$ & $\log_2{n}$ & $n^2$ & $n$ & $n$
 \rule{0pt}{2.6ex}
 \\ 
 \hline
 Ancilla & 0 & 0 & 0 & $n$ & 0 & 0 & 0
 \\ 
 \hline
 Qudit Types & Controls are qudits & Controls are qutrits & Qubits & Qubits & Qubits & Controls are qutrits/qudits & Target is $d=n$-level qudit
  \\ 
  \hline
  Constants &  2 & 13 & 9 & 9 & 9 & 2 & 9
  \\ 
  \hline
  Generalization & $d$-ary & Binary & Binary & Binary & Binary & Ternary/$d$-ary & Binary\\
  \hline
\end{tabular}}

\label{tab:n_controlled}
\end{table*}

As shown in FIG. \ref{8qubit}(a), a multi-controlled Toffoli gate with 7 control qubits and 1 target qubit is considered. FIG. \ref{8qubit}(b) depicts the decomposition of the generalized 8-qubit Toffoli gate as shown in FIG. \ref{8qubit}(a) with the help of the design proposed by Gokhale et al. \cite{gokhale2019asymptotic}. Their circuit temporarily stores information directly in the qutrit $\ket{2}$ state of the controls, so does our approach. However, instead of storing temporary results further with quaternary $\ket{3}$ state, they simply decompose their ternary Toffoli into 13 one-qutrit and two-qutrit gates \cite{Di_2013} as shown in FIG. \ref{toffoli_ternary} \cite{gokhale2019asymptotic}. In our approach, we decompose ternary Toffoli further into three ternary and/or quaternary CNOT gates using $\ket{3}$ state as control, as shown in FIG. \ref{8qubit}(c) on FIG. \ref{8quditdecompose}, three ternary and/or quaternary CNOT gates can be further reduced to two ternary and/or quaternary CNOT gates by using identity rule, as shown in FIG. \ref{8qubit}(d). Thus, our optimization can lead to the reduction of gate-count from 13 to 2 such for a single Toffoli decomposition and our approach can be generalised for qudits also.

Our circuit construction as shown in \ref{8qubit}(c) or \ref{8qubit}(d), as in \cite{gokhale2019asymptotic}, can also be interpreted
as a binary tree of gates. More elaborately, the inputs/outputs are qubits, but we grant inhibition of the $\ket{2}$ and $\ket{3}$ quaternary states in between. The circuit maintains a tree structure and has the property that the intermediate qubit, of each sub-tree as well as root can only be  raised to $\ket{2}$ if all of its seven control leaves were $\ket{1}$. In order to verify this property, we perceive that the qubit $q_4$ can only become $\ket{2}$ if and only if it was originally $\ket{1}$ and qubit $q_6$ was previously $\ket{3}$. At the following level of the tree, we see qubit $q_6$ could have only been $\ket{3}$ if it was previously $\ket{1}$ and both $q_3$ and $q_7$ qubits were $\ket{2}$ before. If any of the controls were not $\ket{1}$, the $\ket{2}$ or $\ket{3}$ states would fail to move to the root of the tree. Hence, the $X$ gate is only carried out if all controls are $\ket{1}$.  The right half of the circuit undergoes computation to get back the controls to their original state. The construction applies more generally to any multi-controlled $U$ gate. 

After each succeeding level of the tree structure, the number of qubits under inspection is reduced by a factor of $\sim2$. This leads to  the circuit depth being logarithmic in $n$, where $n$ is the number of controls. On top of that, each quaternary qudit (termed as ququad) is operated on by a small constant number of three gates, so the total number of gates is optimized. Wang et al. Prior work of \cite{wang} has also mentioned about $n$-qudit Toffoli decomposition. In Table \ref{tab:n_controlled}, we have shown that our approach gives better result in terms of circuit depth and gate cost than \cite{wang}. Further, we have shown an 8-qudit Toffoli decomposition but this can be easily extended to $n$-qudit also, thus our proposed approach is generalized in nature. 

\begin{figure*}[!htb]
 \centering
 \includegraphics[width=185mm,height=10cm]{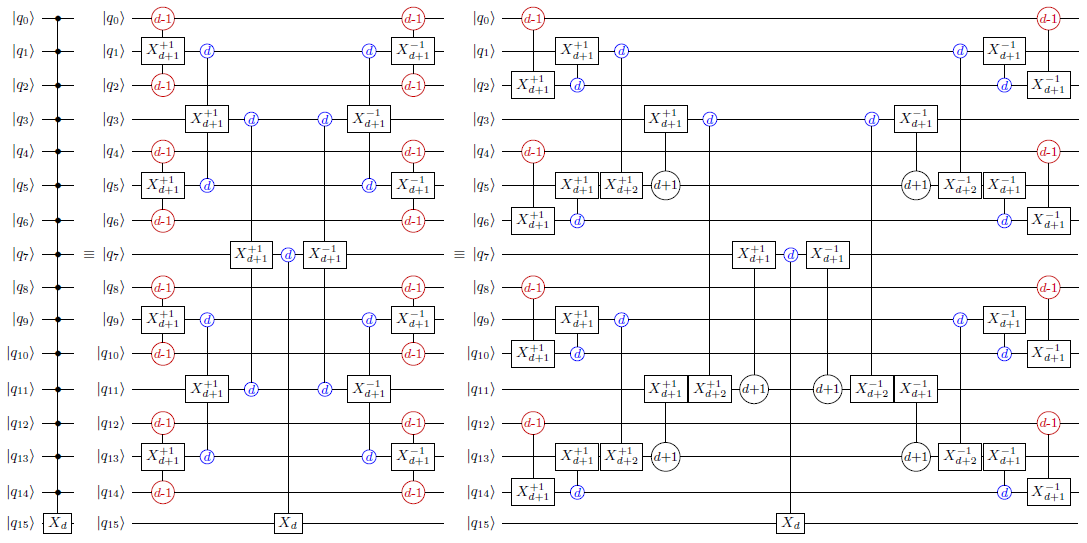}
 \caption{Decomposition of 16-qudit Toffoli Gate.}
 \label{16quditdecompose}
 \end{figure*}

The proposed $n$-qudit Toffoli decomposition is novel not only for its logarithmic depth optimization as compared to \cite{wang}, but also the maximum number of CNOT gates required is $2n - 3$ ($n+1$ number of $d+1$-ary CNOT gates and $n-4$ number of $d+2$-ary CNOT gates), which is less compared to $2n + 1$ needed by the decomposition by Wang et al. \cite{wang}. We have also illustrated an example with 16-qudit Toffoli decomposition elaborately in FIG. \ref{16quditdecompose} for better understanding. Mapping the structure to a binary tree topology helps in establishing the claim for logarithmic depth.

Recall that in a $d$-ary quantum systems, generalized Grover’s algorithm for search over $N$ unstructured database items requires
 $O(\sqrt{N})$ iterations of Grover's operator, where $ N=d^n$ and $d \ge 2$. As discussed earlier, the Grover's operator is the combination of the oracle and the diffusion. However in each iteration, Grover search has multi-controlled Toffoli gate in diffusion operator
with $M=\ceil{\log_d {N}}$
controls \cite{groverd}. In other words, each of the iterations has  $n$-qudit Toffoli gate or ($\log_d N$)-qudits in Grover's diffusion operator as already discussed in FIG. \ref{diff}. The best known Toffoli decomposition in a qudit system \cite{wang},  specifically in ternary quantum systems shows that the depth of the realized circuit is linear, i.e.,  $\log_d N$ or $n$. But, our  decomposition of the $n$-qudit Toffoli gate leads to a reduction of 
the $\log_d N$ factor in Grover’s algorithm to a $\log_2 \log_d N$ factor in each iteration. Hence, our proposed $n$-qudit Toffoli decomposition achieves a logarithmic factor in circuit-depth, i.e., $O(\log_2 \log_d N)$ in the
time complexity of generalised Grover search, compared to the previous works \cite{groverd, hunt2020grovers}, as  shown in Table \ref{tab:grover-d}.

\begin{table}[!htb]
\caption{Comparison of circuit depth, i.e., worst case time complexity for $d$-ary Grover's search,  $d > 2$.}
\centering
\resizebox{6.5cm}{!}{%
\begin{tabular}{|c|c|c|c|}
\hline
  & \textbf{This Work} & Hunt \cite{hunt2020grovers} & Ivanov \cite{groverd}\\ \hline
 Depth & $\log_2{(\log_d{N})}$  & $\log_d{N}$ & $\log_d{N}$  \\
  \hline
\end{tabular}}

\label{tab:grover-d}
\end{table}

We have presented the proposed decomposition of a $d$-ary Toffoli gate, and showed that it is superior to other decomposition in the literature in terms of the depth as well as the number of ancilla qudits required. We have considered every gate to be ideal, and therefore, the success rate of the Grover's algorithm remains same as for any other decomposition of the Toffoli Gate. It has been shown in the literature that if the input state is an entangled state (e.g. GHZ or W state), then the success rate of the algorithm is $(1 - \frac{1}{n})^{n-1}$ where $n$ is the number of qudits \cite{chamoli2007success, Chamoli2007EvolutionOE, 10.1007/s11128-007-0057-2}. However, unlike \cite{chamoli2007success, Chamoli2007EvolutionOE, 10.1007/s11128-007-0057-2}, in this article, we start the algorithm with the equal superposition of all the basis states, and hence the success probability is $\sim 1$ after $\sqrt{N}$ steps.

Next, we address the action of various error models on our decomposition of $n$-qudit Toffoli gate.

\section{Effect of error on proposed decomposition of $n$-qudit Toffoli gate}

Any quantum system is susceptible to different types of errors such as decoherence, noisy gates. For a $d$-ary quantum systems, the gate error scales as $d^2$ and $d^4$ for one and two-qudit gates respectively \cite{gokhale2019asymptotic}. Furthermore, for qubits, the amplitude damping error decays the state $\ket{1}$ to $\ket{0}$ with probability $\lambda_1$. For a $d$-ary system, every state in level $\ket{i} \neq \ket{0}$ has a probability $\lambda_i$ of decaying. In other words, the usage of higher dimensional states penalizes the system with more errors. Nevertheless, the effect of these errors on the used decomposition of Toffoli gate has been studied by Gokhale et al. \cite{gokhale2019asymptotic}. They have shown that although the usage of qutrits leads to increased error, the overall error probability of the decomposition is lower than that for the earlier ones since the number of ancilla qubits and the depth are both reduced. In this section we study the effect of generic gate and relaxation error on our newly proposed decomposition, and show it to be superior to the decomposition of \cite{gokhale2019asymptotic} in terms of error probability.

\subsection{Generic Error Model}
The common quantum error or noise model is for gate and relaxation error \cite{gokhale2019asymptotic}, which can be expressed by the Kraus Operator formalism \cite{chuang}. If the density matrix representation of a (pure) quantum state is $\sigma = \ket{\Psi}{\bra{\Psi}}$, the evolution of this state for any channel is represented as the function $\mathcal{E}(\sigma)$:

\begin{align}
    \mathcal{E}\left(\sigma\right) = \mathcal{E}\left(\ket{\Psi}\bra{\Psi}\right) = \sum_i K_i \sigma K_i^\dagger
\end{align}
where $K_i$ are called the Kraus Operators, and $K_i^\dagger$ is the matrix conjugate-transpose of $K_i$, $\forall$ $i$. Evolution of a state under a noise model can also be represented by the Kraus operator formulation. For example, in the depolarization noise model, the Kraus operators are simply the Pauli matrices.

\subsubsection{Gate Errors}
In a binary quantum system with only one-qubit and two-qubit gates, there are four possible error channels for a one-qubit gate, which can be expressed as products of the two Pauli matrices, a NOT gate, $X = \begin{pmatrix}
0 & 1 \\
1 & 0
\end{pmatrix}$ and a phase gate, $Z = \begin{pmatrix}
1 & 0 \\
0 & -1
\end{pmatrix}$. The possible error channels are: (i) no-error $X^0 Z^0 = I$, (ii) the phase flip which is the product $X^0 Z^1$, (iii) the bit flip which is $X^1 Z^0$ and (iv) the phase+bit flip channel given by $X^1 Z^1$. We can express this one-qubit gate error model in the Kraus operator formalism in the following manner:
\begin{align}
    \mathcal{E}(\sigma) = \sum_{j=0}^{1} \sum_{k=0}^{1} p_{jk} (X^j Z^k) \sigma (X^j Z^k)^{\dagger}
\end{align}
where $p_{jk}$ is the probability of the corresponding Kraus operator.  For a symmetric depolarizing noise model, $p_{jk} = p_{qr}$, $\forall j,k,q,r \in \{0,1\}$, and for an asymmetric noise model \cite{Cafaro_2010} $p_{jk}$ and $p_{qr}$ are in general different for $jk \neq qr$. The total probability of error $p = \displaystyle \sum_{a,b \in \{0,1\}} p_{ab}$.


A noisy gate is modelled as an ideal gate followed by an unwanted Pauli operator \cite{fowler2012surface}. In other words, a one-qubit gate is followed by an unwanted Pauli $\in \{X, Y, Z\}$ with probability $p_x, p_y, p_z$ respectively; and a two-qubit gate is followed by an unwanted Pauli $\in \{I, X, Y, Z\}^{\otimes 2} \setminus \{I,I\}$ with probability $p_i \cdot p_j$, where $i,j \in \{x, y, z\}$. For the sake of convenience, we represent the one-qubit and two-qubit gate error probabilities as $p_1$ and $p_2$ respectively.

In a $d$-ary system, there are $d$ types each of unwanted $X$ and $Z$ Pauli errors that can follow a one-qudit gate \cite{majumdar2020approximate, majumdar2019optimal}. Therefore, there are $d^2 -1$ ways (without considering the identity error) in which an error can occur after a one-qudit gate. If $p_1$ is the probability of a one-qudit Pauli error, then the evolution of the system under noisy one-qudit operations can be represented as in Eq.~\ref{eq:single_gate_err}.

\begin{align}
    \mathcal{E}(\sigma) = (1-(d^2-1)p_1)\sigma + \sum_{jk \in \{0,1\}^d \setminus 0*d} p_{jk} K_{jk} \sigma K_{jk}^{\dagger}
\label{eq:single_gate_err}
\end{align}

where $K_{jk}$ represents the various Pauli operators.

Similarly, for two-qudit gates, an unwanted Pauli operator can occur on each of the two qudits after the gate operation. Therefore, there are $d^4-1$ ways (excluding the identity operation on both the qudits) in which a gate can be noisy. If $p_2$ is the probability of two-qudit gate errors, then the evolution of the system under noisy two-qudit operations is represented as in Eq.~\ref{eq:two_gate_err}.

\begin{equation}
\resizebox{0.97\hsize}{!}{%
    $\mathcal{E}(\sigma) = (1-(d^4-1)p_2)\sigma + \sum_{jklm \in \{0,1\}^d \setminus 0*d} p_{jklm} K_{jklm} \sigma K_{jklm}^{\dagger}$
\label{eq:two_gate_err}}
\end{equation}

where $p_{jklm} = p_{jk} \cdot p_{lm}$. The probability that the density matrix remains error free is independent of whether the underlying depolarizing channel is symmetric or asymmetric. Rather, it depends on the total probability of error.

Our proposed decomposition here  deals with two-qudit gates only on higher dimensional
quantum systems. In general, for the decomposition of a $n$-qudit Toffoli gate, our method uses upto $d+2$ dimension. Therefore, for a $d$ dimensional system, the error in our system scales as $\mathcal{O}(d+2)^4$ as shown in Eq.~\ref{eq:two qudit gate error}. 

\begin{equation}
\resizebox{0.98\hsize}{!}{%
   $ \mathcal{E}(\sigma) = \{1-((d+2)^4-1)p_2\}\sigma + \sum_{\substack{jklm \in \\ \{0,1,2, \dots, d+1\}^4 \setminus 0000}} p_{jklm} K_{jklm} \sigma K_{jklm}^{\dagger}$
\label{eq:two qudit gate error}}
\end{equation}

For example,  there are 16 two-qubit gate error channels for $d=2$, whereas our decomposition requires usage upto $d=4$. Therefore, there are 256 two-ququad gate error channels for our decomposition. In Table~\ref{tab:compgate}, we show the decrease in the probability of no-error for two-qudit gates due to the usage of higher dimensions for $d = 2, 3, 4, 5$. Although our circuit constructions have adopted
higher dimensional gates, our proposed decomposition scales favorably in terms of asymptotically fewer gate errors as our gate count is  asymptotically lower. This is elaborated upon  in \ref{prob_success}.

\begin{table*}[!htb]
    \centering
    \caption{Probability of success of two-qudit gates due to the usage of higher dimensions}
    \resizebox{12.5cm}{!}{%
    \begin{tabular}{|c|c|c|}
    \hline
    
   Dimension $d$ & Probability of success & Probability of success \\
   & without our proposed decomposition & with our proposed decomposition\\
    \hline
    2  & $1 - 15p_2$  & $1 - 255p_2$ \\
    \hline
    3  & $1 - 80p_2$  & $1 - 624p_2$ \\
    \hline
    4  & $1 - 255p_2$  & $1 - 1295p_2$ \\
    \hline
    5  & $1 - 624p_2$  & $1 - 2400p_2$ \\
    \hline
    \end{tabular}}
    \label{tab:compgate}
\end{table*}

\subsubsection{Amplitude Damping Error}
\label{ref:idle errors}

\paragraph{Generalized Amplitude Damping:}

\begin{figure}
    \centering
    \includegraphics[scale=0.45]{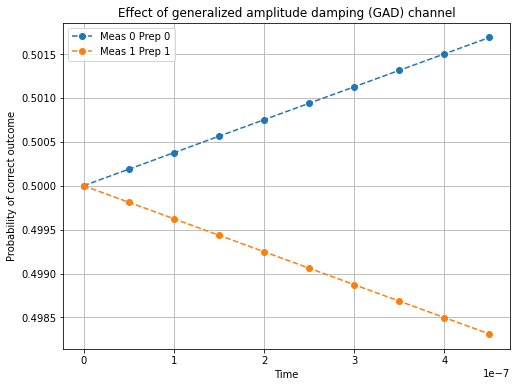}
    \caption{Effect of Generalized Amplitude Damping Channel when $p = 0.5$. Here Meas $i$, Prep $i$, $i \in \{0,1\}$, denotes the probability of measuring a quantum state as $i$ when it was prepared as $i$. We observe that the probability of measuring $\ket{1}$ decreases gradually with time, while that of $\ket{0}$ increases}
    \label{fig:GAD}
\end{figure}

A qubit, when kept idle, can change its state spontaneously. It can (i) absorb some energy from the environment, and move to a higher energy state, or (ii) release energy spontaneously and move from a higher to a lower energy state. In a more generalized scenario, a qubit can both absorb energy with probability $p$ and release energy with probability $1-p$. This error is termed as \emph{Generalized Amplitude Damping (GAD)} \cite{Cafaro_2014}. For qubits, this noise channel is characterized by the Kraus Operators
\begin{equation*}
    K_0 = \sqrt{p}\begin{pmatrix}
        1 & 0 \\
        0 & \sqrt{1-\lambda}
    \end{pmatrix} \quad K_1 = \sqrt{p}\begin{pmatrix}
        0 & \sqrt{\lambda} \\
        0 & 0 \quad 
    \end{pmatrix}
\end{equation*}

\begin{equation*}
    K_2 = \sqrt{1-p}\begin{pmatrix}
        \sqrt{1-\lambda} & 0 \\
        0 & 1
    \end{pmatrix} \quad K_3 = \sqrt{1-p}\begin{pmatrix}
        0 & 0 \\
        \sqrt{\lambda} & 0
    \end{pmatrix}
\end{equation*}
where $\lambda \propto exp(-t/T1)$, $T_1$ being termed as relaxation time. Note that the time duration, $t$, in general depends on the depth of the circuit. Here two of the Kraus Operators govern the change of state from $\ket{0} \rightarrow \ket{1}$, and the other two vice-versa. For a fixed value of $T_1$, this decay is governed by the values of $p$ and $t$. For example, when $p = 0.5$, the initial  state has an equal probability of being in the states $\ket{0}$ or $\ket{1}$. The action of the GAD Kraus operators on this state is to increase the probability of obtaining $\ket{0}$, while decreasing that of $\ket{1}$. In FIG.~\ref{fig:GAD} we show the effect of GAD Kraus operators, when $p = 0.5$.

In FIG.~\ref{fig:GAD}, $T_1$ is taken to be $133 \mu s$, which is the average $T_1$ time of \emph{ibmq\_kolkata}. The time duration is varied from $t = 0$ to $450 ns$, where $445 ns$ is the average duration of a CNOT gate in the same device. We want to emphasize here, that our methodology for multi-controlled Toffoli decomposition achieves logarithmic depth (i.e., the time duration lowers from $\mathcal{O}(t)$ to $\mathcal{O}(\log_2 t)$. In general, execution of higher dimensional quantum circuits is not yet feasible in IBM Quantum Devices. Some studies using Pulse simulation has been made to use one or more higher dimensions (we elaborate on this later). However, although the intensity of noise may vary with dimension, the effect of noise remains similar. Therefore, we give an essence of the flavour of GAD in FIG.~\ref{fig:GAD} for qubit systems only. It readily shows the reduction in the effect of GAD when the time reduces from $\simeq 450 ns$ to $\simeq 10 ns$, which is roughly the logarithm of the former.

Generalized Amplitude Damping is not always notable in quantum computers, since change from lower to higher energy state is not spontaneous, and requires external driving energy. On the other hand, spontaneous relaxation from higher to lower energy state is a major bane in modern quantum devices. In the following subsection, we look deeper into this spontaneous relaxation. Henceforth, the terms \emph{idle error}, \emph{amplitude damping} would both imply this spontaneous relaxation only.

\paragraph{Spontaneous relaxation/Idle error:}

Idle errors usually concentrate on the relaxation from higher to lower energy states in quantum devices. This is also known as amplitude damping. This noise channel irreversibly takes qudits to lower states. For qubits, the only amplitude damping channel is from $\ket{1}$ to $\ket{0}$, and we denote this damping probability as $\lambda_1$. For qubits, the Kraus operators for amplitude damping are:
\begin{align} K_0 = \begin{pmatrix} 1 & 0 \\ 0 & \sqrt{1 - \lambda_1} \end{pmatrix} \text{\quad and \quad} K_1 = \begin{pmatrix} 0 & \sqrt{\lambda_1} \\ 0 & 0 \end{pmatrix}
\end{align}

 For qudits, we also model damping from $\ket{d-1}$ to $\ket{0}$, which occurs with probability $\lambda_{d-1}$. For qudits, the Kraus operator for amplitude damping can be modeled as:

$$
K_0 = \begin{pmatrix} 1 & 0 & 0 & \ldots & 0\\ 0 & \sqrt{1-\lambda_1} & 0 & \ldots & 0 \\ 0 & 0 & \sqrt{1 - \lambda_2} & \ldots & 0 \\ \vdots & \vdots & \vdots & \ddots & \vdots \\ 0 & 0 & 0 & \ldots & \sqrt{1 - \lambda_{d-1}} \end{pmatrix}
\text{, }
$$
$$
K_1 = \begin{pmatrix} 0 & \sqrt{\lambda_1} & 0 & \ldots & 0\\ 0 & 0 & 0 & \ldots & 0\\ 0 & 0 & 0 & \ldots & 0 \\ \vdots & \vdots & \vdots & \ddots & \vdots \\ 0 & 0 & 0 & \ldots & 0 \end{pmatrix}
\text{,}
$$
\begin{align}
\dots
K_{d-1} = \begin{pmatrix} 0 & 0 & 0 & \ldots & \sqrt{\lambda_{d-1}}\\ 0 & 0 & 0 & \ldots & 0\\ 0 & 0 & 0 & \ldots & 0 \\ \vdots & \vdots & \vdots & \ddots & \vdots \\ 0 & 0 & 0 & \ldots & 0 \end{pmatrix}
\end{align}

In each Kraus Operator $K_i$, the value of $\lambda_i \propto exp(-t/T_{1_i})$, where $t$ is the duration of the computation, and $T_{1_i}$ are the relaxation time. Note that in the previous subsection, we have used the well known terminology of $T_1$ for qubit systems instead of $T_{1_1}$. However, in this subsection we shall use $T_{1_i}$ for any finite-dimensional quantum system to avoid confusion. We have qubit quantum devices, where $T_{1_1} \simeq 100 \mu s$ in some higher end IBM Quantum Devices \cite{ibmquantum}. However, due to the lack of qudit quantum computers, we do not have explicit values of other $T_{1_i}$'s except $30 \mu s$ for qutrit ($T_{1_2}$) and quaquad ($T_{1_3}$) quantum devices \cite{https://doi.org/10.48550/arxiv.2203.07369}. Nevertheless, the time duration depends on the circuit depth. Hence, idle errors are reduced by decreasing the circuit \textit{depth}. Recall that the $n$-qudit Toffoli decomposition of \cite{wang} requires circuit depth of $n$, so if $\Delta t$ is the time duration of each quantum operation, the total time required $n\cdot \Delta t$. On the contrary, our proposed decomposition requires only $log_2 n$ circuit depth, so the duration of total computation is $log_2 n \Delta t$. Hence, the decoherence due to our proposed decomposition is  $\mathcal{O}(exp(-log_2 n \Delta t/T_{1_i}))$ which is notably lower than $\mathcal{O}(exp(-n \Delta t/T_{1_i}))$ for the previous decomposition.

\subsection{Comparative analysis of success probability}\label{prob_success}

In \cite{gokhale2019asymptotic}, the authors first proposed the usage of higher dimension for efficient decomposition of Toffoli gates. However, in that article, they restricted themselves for decomposition of binary multi-controlled Toffoli gates using up to three dimensional quantum systems only. Therefore, it is not possible to compare our $d$-dimensional Toffoli decomposition with that result. However, in this section, we study the probability of success in the decomposed circuit of an $n$-qubit Toffoli gate using the method in \cite{gokhale2019asymptotic}, and our proposed method in this article. Note that, while the decomposition of \cite{gokhale2019asymptotic} requires only ternary gates, our proposed decomposition requires ternary and quaternary gates.

The complexity of decomposition of an $n$-qubit Toffoli gate in terms of the number of gates and depth of the circuit for the method in \cite{gokhale2019asymptotic} and our proposed one are depicted in Table~\ref{tab:resource}.

\begin{table*}[htb]
    \centering
    \caption{Number of gates and depth of the decomposed circuit for an $n$-qubit Toffoli gate}
    \begin{tabular}{|c|c|c|c|c|}
    \hline
         & \# one-qutrit gates & \# two-qutrit gates & \# two-ququad gates & depth \\
         \hline
        Decomposition of \cite{gokhale2019asymptotic} & $7(n-2)+1$ & $6(n-2)$ & 0 & $26 \lceil \log_2 n \rceil + 1$\\
        \hline
        Proposed Decomposition & 0 & $n + 1$ & $n - 4$ & $ 4 \lceil \log_2 n \rceil$\\
        \hline
    \end{tabular}
    \label{tab:resource}
\end{table*}

The gates used in quantum circuits often have small errors which can be modelled as an ideal gate followed by an unwanted Pauli operator as mentioned in earlier subsections. However, in this comparison, we follow the approach used in \cite{majumdar2021optimizing}, and instead of comparing the probability of small errors in the circuit, we compare  without loss of generality, the probability that the circuit remains error-free (probability of success) for the decomposition  in \cite{gokhale2019asymptotic} and our proposed decomposition.

For any decomposition, the generalized formula for probability of success ($P_{success}$) is the product that the individual components does not fail. In other words,
\begin{align*}
    P_{success} &=& \Pi_{gates} ({(P_{success~of~ gate})}^{\#~gates}
     \\
    & & \times \; e^{-(depth/T_1)}), 
\end{align*}
where the product in the first term is over all the types of gates used in the decomposition (one-qutrit, two-qutrit, two-ququad), and the second term is the probability of no relaxation error. Note that when a particular type of gate is not used in a decomposition, the corresponding term attains a value of $1$ due to having a zero in the power. For example, our proposed decomposition does not require any one-qutrit gate, and hence the contribution of that term in the product is 1.

Current quantum devices are mostly binary, and the probabilities of one-qubit and two-qubit gates in the IBMQ Quantum Devices are in the range of $10^{-4}$ and $10^{-2}$ respectively \cite{ibmquantum}. Moreover, the time $T_{1_1}$  of most of the IBM Quantum Devices are in the range of $100 \mu s$.  However, in \cite{https://doi.org/10.48550/arxiv.2203.07369}, the authors experimentally showed that the value of $T_{1_2}$ and $T_{1_3}$ for each ternary and quaternary gate is $30 \mu s$, which we have also assumed  for our experiment. We assume that the probability of error of each two-qutrit and two-ququad gate is $10^{-2}$, that of one-qutrit gate is $10^{-4}$, and the time $T_{1_2}$ and $T_{1_3}$ is $30 \mu s$ for our simulation.

\begin{figure}
    \centering
    \includegraphics[scale=0.48]{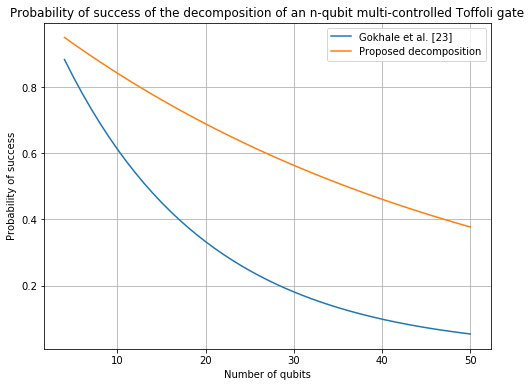}
    \caption{Probability of success for the decomposition of an $n$-qubit Toffoli gate using our proposed method versus the method in \cite{gokhale2019asymptotic}}
    \label{fig:success}
\end{figure}

In FIG.~\ref{fig:success}, we show the probability of success for the multi-controlled Toffoli gate decomposition using the method of \cite{gokhale2019asymptotic} (which we label as Gokhale et al.) and our proposed method. We observe that our proposed method has a significantly lower error rate than the decomposition in \cite{gokhale2019asymptotic}. This can be attributed to the lower number of gates and lower depth of our decomposition. Although our decomposition uses a few ququad gates, which have a higher error probability due to the curse of dimensionality (Table~\ref{tab:compgate}), the overall significant decrease in the gate count and the depth makes our method superior. In fact, for $n = 50$, our decomposition has a probability of success of $\simeq 0.4$, whereas that of \cite{gokhale2019asymptotic} has a probability of success of $\simeq 0.05$. Therefore, we obtain a percentage decrease in the probability of error by $\simeq 40\%$ for $n = 50$.

\section{Conclusion}\label{sec5}
In this work, we have proposed a novel approach to decompose a generalized $n$-qudit Toffoli gate into two-qudit gates with logarithmic depth without using any ancilla qudit. We have shown an instance of $8$-qudit Toffoli gate decomposition to establish the logarithmic depth as an example in Section \ref{section3a}. We have given a comparative study to establish that our approach is better than the existing state-of-the-art ones  in Table \ref{tab:n_controlled} in Section \ref{section3b}, where it has been shown that the constant factor of gate-count from 13
to 2 for single Toffoli decomposition for $n$-qubit Toffoli gate. We have also shown that Grover's algorithm can be implemented in any $d$-ary quantum system with the proposed $n$-qudit Toffoli gate to get the advantage of optimized logarithmic depth as compared to earlier approaches as shown in Table \ref{tab:grover-d} in Section \ref{section3b}. Using our novel proposed decomposition of $n$-qudit Toffoli gate, any quantum algorithm can be optimized that employs generalized Toffoli gate. Finally, we have studied the effect of different error models on this decomposition technique. Our study shows that the few  gates in higher dimensional quantum systems which are used in the proposed decomposition, are prone to more errors. Nevertheless, as we have obtained asymptotically improved gate count and circuit depth, leading to low total error probability, the gates can operate with high fidelity as compared to state-of-the art works in FIG. \ref{fig:success} of Section \ref{prob_success}.

\begin{acknowledgments}
The first author acknowledges the support by the  Grant No. 09/028(0987)/2016-EMR-I from CSIR, Govt. of India.
\end{acknowledgments}


\bibliography{apssamp}

\appendix

\section{Simulation Results}

The simulation result for 8-qubit Toffoli gate, FIG. \ref{8qubit}(c), is given below. We have shown that only for the input state $\ket{11111110}$, the output state changes to $\ket{11111111}$, which is highlighted in Table \ref{simu}, otherwise there is no change of output states for corresponding input states. Similarly, if we initialize $q_8$ of FIG. \ref{8qubit}(c) with $\ket{1}$, then the output state changes to $\ket{11111110}$ for the input state $\ket{11111111}$. There is no change of output states for other corresponding input states as well. The simulation is carried out on Google Colab platform \cite{Bisong2019} and the code is available at \href{https://github.com/amitsaha2806/N-Qudit-Toffoli-Decomposition}{https://github.com/N-Qudit-Toffoli-Decomposition}.

\begin{table*}[!htb]
    \centering
    \caption{Simulation result for 8-qubit Toffoli shown in Figure \ref{8qubit}.}
    \resizebox{\textwidth}{!}{%
    \begin{tabular}{|c|c|c|c|c|c|c|c|c|}
    \hline
    Input State & After $1^{st}$ time cycle & After $2^{nd}$ time cycle & After $3^{rd}$ time cycle & After $4^{th}$ time cycle & After $5^{th}$ time cycle & After $6^{th}$ time cycle & After $7^{th}$ time cycle & Output State after Mirror 
\\
\hline

$\ket{00000000}$ & $\ket{00000000}$ & $\ket{00000000}$ & $\ket{00000000}$ & $\ket{00000000}$ & $\ket{00000000}$ & $\ket{00000000}$ & $\ket{00000000}$ & $\ket{00000000}$
\\
\hline

$\ket{00000010}$ & $\ket{00000010}$ & $\ket{00000010}$ & $\ket{00000010}$ & $\ket{00000010}$ & $\ket{00000010}$ & $\ket{00000010}$ & $\ket{00000010}$ & $\ket{00000010}$
\\
\hline

$\ket{00000100}$ & $\ket{00000100}$ & $\ket{00000100}$ & $\ket{00000100}$ & $\ket{00000100}$ & $\ket{00000100}$ & $\ket{00000100}$ & $\ket{00000100}$ & $\ket{00000100}$
\\
\hline

$\ket{00000110}$ & $\ket{00000110}$ & $\ket{00000110}$ & $\ket{00000110}$ & $\ket{00000110}$ & $\ket{00000110}$ & $\ket{00000110}$ & $\ket{00000110}$ & $\ket{00000110}$
\\
\hline

$\ket{00001000}$ & $\ket{00001010}$ & $\ket{00001010}$ & $\ket{00001000}$ & $\ket{00001000}$ & $\ket{00001000}$ & $\ket{00001000}$ & $\ket{00001000}$ & $\ket{00001000}$
\\
\hline

$\ket{00001010}$ & $\ket{00001020}$ & $\ket{00001120}$ & $\ket{00001110}$ & $\ket{00001110}$ & $\ket{00001110}$ & $\ket{00001110}$ & $\ket{00001110}$ & $\ket{00001010}$
\\
\hline

$\ket{00001100}$ & $\ket{00001110}$ & $\ket{00001110}$ & $\ket{00001100}$ & $\ket{00001100}$ & $\ket{00001100}$ & $\ket{00001100}$ & $\ket{00001100}$ & $\ket{00001100}$
\\
\hline

$\ket{00001110}$ & $\ket{00001120}$ & $\ket{00001220}$ & $\ket{00001210}$ & $\ket{00001210}$ & $\ket{00001210}$ & $\ket{00001210}$ & $\ket{00001210}$ & $\ket{00001110}$
\\
\hline

$\ket{00010000}$ & $\ket{00010000}$ & $\ket{00010000}$ & $\ket{00010000}$ & $\ket{00010000}$ & $\ket{00010000}$ & $\ket{00010000}$ & $\ket{00010000}$ & $\ket{00010000}$
\\
\hline

$\ket{00010010}$ & $\ket{00010010}$ & $\ket{00010010}$ & $\ket{00010010}$ & $\ket{00010010}$ & $\ket{00010010}$ & $\ket{00010010}$ & $\ket{00010010}$ & $\ket{00010010}$
\\
\hline

$\ket{00010100}$ & $\ket{00010100}$ & $\ket{00010100}$ & $\ket{00010100}$ & $\ket{00010100}$ & $\ket{00010100}$ & $\ket{00010100}$ & $\ket{00010100}$ & $\ket{00010100}$
\\
\hline

$\ket{00010110}$ & $\ket{00010110}$ & $\ket{00010110}$ & $\ket{00010110}$ & $\ket{00010110}$ & $\ket{00010110}$ & $\ket{00010110}$ & $\ket{00010110}$ & $\ket{00010110}$
\\
\hline

$\ket{00011000}$ & $\ket{00011010}$ & $\ket{00011010}$ & $\ket{00011000}$ & $\ket{00011000}$ & $\ket{00011000}$ & $\ket{00011000}$ & $\ket{00011000}$ & $\ket{00011000}$
\\
\hline

$\ket{00011010}$ & $\ket{00011020}$ & $\ket{00011120}$ & $\ket{00011110}$ & $\ket{00011110}$ & $\ket{00011110}$ & $\ket{00011110}$ & $\ket{00011110}$ & $\ket{00011010}$
\\
\hline

$\ket{00011100}$ & $\ket{00011110}$ & $\ket{00011110}$ & $\ket{00011100}$ & $\ket{00011100}$ & $\ket{00011100}$ & $\ket{00011100}$ & $\ket{00011100}$ & $\ket{00011100}$
\\
\hline

$\ket{00011110}$ & $\ket{00011120}$ & $\ket{00011220}$ & $\ket{00011210}$ & $\ket{00011210}$ & $\ket{00011210}$ & $\ket{00011210}$ & $\ket{00011210}$ & $\ket{00011110}$
\\
\hline

$\ket{00100000}$ & $\ket{00100000}$ & $\ket{00100000}$ & $\ket{00100000}$ & $\ket{00100000}$ & $\ket{00100000}$ & $\ket{00100000}$ & $\ket{00100000}$ & $\ket{00100000}$
\\
\hline

$\ket{00100010}$ & $\ket{00100010}$ & $\ket{00100010}$ & $\ket{00100010}$ & $\ket{00100010}$ & $\ket{00100010}$ & $\ket{00100010}$ & $\ket{00100010}$ & $\ket{00100010}$
\\
\hline

$\ket{00100100}$ & $\ket{00100100}$ & $\ket{00100100}$ & $\ket{00100100}$ & $\ket{00100100}$ & $\ket{00100100}$ & $\ket{00100100}$ & $\ket{00100100}$ & $\ket{00100100}$
\\
\hline

$\ket{00100110}$ & $\ket{00100110}$ & $\ket{00100110}$ & $\ket{00100110}$ & $\ket{00100110}$ & $\ket{00100110}$ & $\ket{00100110}$ & $\ket{00100110}$ & $\ket{00100110}$
\\
\hline

$\ket{00101000}$ & $\ket{00101010}$ & $\ket{00101010}$ & $\ket{00101000}$ & $\ket{00101000}$ & $\ket{00101000}$ & $\ket{00101000}$ & $\ket{00101000}$ & $\ket{00101000}$
\\
\hline

$\ket{00101010}$ & $\ket{00101020}$ & $\ket{00101120}$ & $\ket{00101110}$ & $\ket{00101110}$ & $\ket{00101110}$ & $\ket{00101110}$ & $\ket{00101110}$ & $\ket{00101010}$
\\
\hline

$\ket{00101100}$ & $\ket{00101110}$ & $\ket{00101110}$ & $\ket{00101100}$ & $\ket{00101100}$ & $\ket{00101100}$ & $\ket{00101100}$ & $\ket{00101100}$ & $\ket{00101100}$
\\
\hline

$\ket{00101110}$ & $\ket{00101120}$ & $\ket{00101220}$ & $\ket{00101210}$ & $\ket{00101210}$ & $\ket{00101210}$ & $\ket{00101210}$ & $\ket{00101210}$ & $\ket{00101110}$
\\
\hline

$\ket{00110000}$ & $\ket{00110000}$ & $\ket{00110000}$ & $\ket{00110000}$ & $\ket{00110000}$ & $\ket{00110000}$ & $\ket{00110000}$ & $\ket{00110000}$ & $\ket{00110000}$
\\
\hline

$\ket{00110010}$ & $\ket{00110010}$ & $\ket{00110010}$ & $\ket{00110010}$ & $\ket{00110010}$ & $\ket{00110010}$ & $\ket{00110010}$ & $\ket{00110010}$ & $\ket{00110010}$
\\
\hline

$\ket{00110100}$ & $\ket{00110100}$ & $\ket{00110100}$ & $\ket{00110100}$ & $\ket{00110100}$ & $\ket{00110100}$ & $\ket{00110100}$ & $\ket{00110100}$ & $\ket{00110100}$
\\
\hline

$\ket{00110110}$ & 
$\ket{00110110}$ & 
$\ket{00110110}$ & 
$\ket{00110110}$ & 
$\ket{00110110}$ & 
$\ket{00110110}$ & 
$\ket{00110110}$ & 
$\ket{00110110}$ & 
$\ket{00110110}$
\\
\hline

$\ket{00111000}$ & $\ket{00111010}$ & $\ket{00111010}$ & $\ket{00111000}$ & $\ket{00111000}$ & $\ket{00111000}$ & $\ket{00111000}$ & $\ket{00111000}$ & $\ket{00111000}$
\\
\hline

$\ket{00111010}$ & $\ket{00111020}$ & $\ket{00111120}$ & $\ket{00111110}$ & $\ket{00111110}$ & $\ket{00111110}$ & $\ket{00111110}$ & $\ket{00111110}$ & $\ket{00111010}$
\\
\hline

$\ket{00111100}$ & $\ket{00111110}$ & $\ket{00111110}$ & $\ket{00111100}$ & $\ket{00111100}$ & $\ket{00111100}$ & $\ket{00111100}$ & $\ket{00111100}$ & $\ket{00111100}$
\\
\hline

$\ket{00111110}$ & $\ket{00111120}$ & $\ket{00111220}$ & $\ket{00111210}$ & $\ket{00111210}$ &
$\ket{00111210}$ &
$\ket{00111210}$ &
$\ket{00111210}$ &
$\ket{00111110}$
\\

\hline

$\ket{01000000}$ & $\ket{01000000}$ & $\ket{01000000}$ & $\ket{01000000}$ & $\ket{01000000}$ & $\ket{01000000}$ & $\ket{01000000}$ & $\ket{01000000}$ & $\ket{01000000}$
\\
\hline

$\ket{01000010}$ & $\ket{01000010}$ & $\ket{01000010}$ & $\ket{01000010}$ & $\ket{01000010}$ & $\ket{01000010}$ & $\ket{01000010}$ & $\ket{01000010}$ &  $\ket{01000010}$
\\
\hline

$\ket{01000100}$ & $\ket{01000100}$ &
$\ket{01000100}$ &
$\ket{01000100}$ & $\ket{01000100}$ & $\ket{01000100}$ & $\ket{01000100}$ & $\ket{01000100}$ & $\ket{01000100}$
\\
\hline

$\ket{01000110}$ & $\ket{01000110}$ & $\ket{01000110}$ & $\ket{01000110}$ & $\ket{01000110}$ & $\ket{01000110}$ & $\ket{01000110}$ & $\ket{01000110}$ & $\ket{01000110}$
\\
\hline

$\ket{01001000}$ & $\ket{01001010}$ & $\ket{01001010}$ & $\ket{01001000}$ & $\ket{01001000}$ & $\ket{01001000}$ & $\ket{01001000}$ & $\ket{01001000}$ & $\ket{01001000}$
\\
\hline

$\ket{01001010}$ & $\ket{01001020}$ & $\ket{01001120}$ & $\ket{01001110}$ & $\ket{01001110}$ & $\ket{01001110}$ & $\ket{01001110}$ & $\ket{01001110}$ & $\ket{01001010}$
\\
\hline

$\ket{01001100}$ & $\ket{01001110}$ & $\ket{01001110}$ & $\ket{01001100}$ & $\ket{01001100}$ & $\ket{01001100}$ & $\ket{01001100}$ & $\ket{01001100}$ & $\ket{01001100}$
\\
\hline

$\ket{01001110}$ & $\ket{01001120}$ & $\ket{01001220}$ & $\ket{01001210}$ & $\ket{01001210}$ & $\ket{01001210}$ & $\ket{01001210}$ & $\ket{01001210}$ & $\ket{01001110}$
\\
\hline

$\ket{01010000}$ & $\ket{01010000}$ & $\ket{01010000}$ & $\ket{01010000}$ & $\ket{01010000}$ & $\ket{01010000}$ & $\ket{01010000}$ & $\ket{01010000}$ & $\ket{01010000}$
\\
\hline

$\ket{01010010}$ & $\ket{01010010}$ & $\ket{01010010}$ & $\ket{01010010}$ & $\ket{01010010}$ & $\ket{01010010}$ & $\ket{01010010}$ & $\ket{01010010}$ & $\ket{01010010}$
\\
\hline

$\ket{01010100}$ & $\ket{01010100}$ & $\ket{01010100}$ & $\ket{01010100}$ & $\ket{01010100}$ & $\ket{01010100}$ & $\ket{01010100}$ & $\ket{01010100}$ & $\ket{01010100}$
\\
\hline

$\ket{01010110}$ & $\ket{01010110}$ & $\ket{01010110}$ & $\ket{01010110}$ & $\ket{01010110}$ & $\ket{01010110}$ & $\ket{01010110}$ & $\ket{01010110}$ & $\ket{01010110}$
\\
\hline

$\ket{01011000}$ & $\ket{01011010}$ & $\ket{01011010}$ & $\ket{01011000}$ & $\ket{01011000}$ & $\ket{01011000}$ & $\ket{01011000}$ & $\ket{01011000}$ & $\ket{01011000}$
\\
\hline

$\ket{01011010}$ & $\ket{01011020}$ & $\ket{01011120}$ & $\ket{01011110}$ & $\ket{01011110}$ & $\ket{01011110}$ & $\ket{01011110}$ & $\ket{01011110}$ & $\ket{01011010}$
\\
\hline

$\ket{01011100}$ & $\ket{01011110}$ & $\ket{01011110}$ & $\ket{01011100}$ & $\ket{01011100}$ & $\ket{01011100}$ & $\ket{01011100}$ & $\ket{01011100}$ & $\ket{01011100}$
\\
\hline

$\ket{01011110}$ & $\ket{01011120}$ & $\ket{01011220}$ & $\ket{01011210}$ & $\ket{01011210}$ & $\ket{01011210}$ & $\ket{01011210}$ & $\ket{01011210}$ & $\ket{01011110}$
\\

\hline

$\ket{01100000}$ & $\ket{01100000}$ & $\ket{01100000}$ & $\ket{01100000}$ & $\ket{01100000}$ & $\ket{01100000}$ & $\ket{01100000}$ & $\ket{01100000}$ & $\ket{01100000}$
\\
\hline

$\ket{01100010}$ & $\ket{01100010}$ & $\ket{01100010}$ & $\ket{01100010}$ & $\ket{01100010}$ & $\ket{01100010}$ & $\ket{01100010}$ & $\ket{01100010}$ & $\ket{01100010}$
\\
\hline

$\ket{01100100}$ & $\ket{01100100}$ & $\ket{01100100}$ & $\ket{01100100}$ & $\ket{01100100}$ & $\ket{01100100}$ & $\ket{01100100}$ & $\ket{01100100}$ & $\ket{01100100}$
\\
\hline

$\ket{01100110}$ & $\ket{01100110}$ & $\ket{01100110}$ & $\ket{01100110}$ & $\ket{01100110}$ & $\ket{01100110}$ & $\ket{01100110}$ & $\ket{01100110}$ & $\ket{01100110}$
\\
\hline

$\ket{01101000}$ & $\ket{01101010}$ & $\ket{01101010}$ & $\ket{01101000}$ & $\ket{01101000}$ & $\ket{01101000}$ & $\ket{01101000}$ & $\ket{01101000}$ & $\ket{01101000}$
\\
\hline

$\ket{01101010}$ & $\ket{01101020}$ & $\ket{01101120}$ & $\ket{01101110}$ & $\ket{01101110}$ & $\ket{01101110}$ & $\ket{01101110}$ & $\ket{01101110}$ & $\ket{01101010}$
\\
\hline

$\ket{01101100}$ & $\ket{01101110}$ & $\ket{01101110}$ & $\ket{01101100}$ & $\ket{01101100}$ & $\ket{01101100}$ & $\ket{01101100}$ & $\ket{01101100}$ & $\ket{01101100}$
\\
\hline

$\ket{01101110}$ & $\ket{01101120}$ & $\ket{01101220}$ & $\ket{01101210}$ & $\ket{01101210}$ & $\ket{01101210}$ & $\ket{01101210}$ & $\ket{01101210}$ & $\ket{01101110}$
\\
\hline

$\ket{01110000}$ & $\ket{01110000}$ & $\ket{01110000}$ & $\ket{01110000}$ & $\ket{01110000}$ & $\ket{01110000}$ & $\ket{01110000}$ & $\ket{01110000}$ & $\ket{01110000}$
\\
\hline

$\ket{01110010}$ & $\ket{01110010}$ & $\ket{01110010}$ & $\ket{01110010}$ & $\ket{01110010}$ & $\ket{01110010}$ & $\ket{01110010}$ & $\ket{01110010}$ & $\ket{01110010}$
\\
\hline

$\ket{01110100}$ & $\ket{01110100}$ & $\ket{01110100}$ & $\ket{01110100}$ & $\ket{01110100}$ & $\ket{01110100}$ & $\ket{01110100}$ & $\ket{01110100}$ & $\ket{01110100}$
\\
\hline

$\ket{01110110}$ & $\ket{01110110}$ & $\ket{01110110}$ & $\ket{01110110}$ & $\ket{01110110}$ & $\ket{01110110}$ & $\ket{01110110}$ & $\ket{01110110}$ & $\ket{01110110}$
\\
\hline

$\ket{01111000}$ & $\ket{01111010}$ & $\ket{01111010}$ & $\ket{01111000}$ & $\ket{01111000}$ & $\ket{01111000}$ & $\ket{01111000}$ & $\ket{01111000}$ & $\ket{01111000}$
\\
\hline

$\ket{01111010}$ & $\ket{01111020}$ & $\ket{01111120}$ & $\ket{01111110}$ & $\ket{01111110}$ & $\ket{01111110}$ & $\ket{01111110}$ & $\ket{01111110}$ & $\ket{01111010}$
\\
\hline

$\ket{01111100}$ & $\ket{01111110}$ & $\ket{01111110}$ & $\ket{01111100}$ & $\ket{01111100}$ & $\ket{01111100}$ & $\ket{01111100}$ & $\ket{01111100}$ & $\ket{01111100}$
\\
\hline

$\ket{01111110}$ & $\ket{01111120}$ & $\ket{01111220}$ & $\ket{01111210}$ & $\ket{01111210}$ & $\ket{01111210}$ & $\ket{01111210}$ & $\ket{01111210}$ & $\ket{01111110}$
\\

\hline

$\ket{10000000}$ & $\ket{10100000}$ & $\ket{10100000}$ & $\ket{10000000}$ & $\ket{10000000}$ & $\ket{10000000}$ & $\ket{10000000}$ & $\ket{10000000}$ & $\ket{10000000}$
\\
\hline

$\ket{10000010}$ & $\ket{10100010}$ & $\ket{10100010}$ & $\ket{10000010}$ & $\ket{10000010}$ & $\ket{10000010}$ & $\ket{10000010}$ & $\ket{10000010}$ & $\ket{10000010}$
\\
\hline

$\ket{10000100}$ & $\ket{10100100}$ & $\ket{10100100}$ & $\ket{10000100}$ & $\ket{10000100}$ & $\ket{10000100}$ & $\ket{10000100}$ & $\ket{10000100}$ & $\ket{10000100}$
\\

\hline

$\ket{10000110}$ & $\ket{10100110}$ & $\ket{10100110}$ & $\ket{10000110}$ & $\ket{10000110}$ & $\ket{10000110}$ & $\ket{10000110}$ & $\ket{10000110}$ & $\ket{10000110}$
\\
\hline

$\ket{10001000}$ & $\ket{10101010}$ & $\ket{10101010}$ & $\ket{10001000}$ & $\ket{10001000}$ & $\ket{10001000}$ & $\ket{10001000}$ & $\ket{10001000}$ & $\ket{10001000}$
\\
\hline

$\ket{10001010}$ & $\ket{10101020}$ & $\ket{10101120}$ & $\ket{10001110}$ & $\ket{10001110}$ & $\ket{10001110}$ & $\ket{10001110}$ & $\ket{10001110}$ & $\ket{10001010}$
\\
\hline

$\ket{10001100}$ & $\ket{10101110}$ & $\ket{10101110}$ & $\ket{10001100}$ & $\ket{10001100}$ & $\ket{10001100}$ & $\ket{10001100}$ & $\ket{10001100}$ & $\ket{10001100}$
\\

\hline

$\ket{10001110}$ & $\ket{10101120}$ & $\ket{10101220}$ & $\ket{10001210}$ & $\ket{10001210}$ & $\ket{10001210}$ & $\ket{10001210}$ & $\ket{10001210}$ & $\ket{10001110}$
\\
\hline

$\ket{10010000}$ & $\ket{10110000}$ & $\ket{10110000}$ & $\ket{10010000}$ & $\ket{10010000}$ & $\ket{10010000}$ & $\ket{10010000}$ & $\ket{10010000}$ & $\ket{10010000}$
\\
\hline

$\ket{10010010}$ & $\ket{10110010}$ & $\ket{10110010}$ & $\ket{10010010}$ & $\ket{10010010}$ & $\ket{10010010}$ & $\ket{10010010}$ & $\ket{10010010}$ & $\ket{10010010}$
\\

\hline

$\ket{10010100}$ & $\ket{10110100}$ & $\ket{10110100}$ & $\ket{10010100}$ & $\ket{10010100}$ & $\ket{10010100}$ & $\ket{10010100}$ & $\ket{10010100}$ & $\ket{10010100}$
\\
\hline

$\ket{10010110}$ & $\ket{10110110}$ & $\ket{10110110}$ & $\ket{10010110}$ & $\ket{10010110}$ & $\ket{10010110}$ & $\ket{10010110}$ & $\ket{10010110}$ & $\ket{10010110}$
\\
\hline

$\ket{10011000}$ & $\ket{10111010}$ & $\ket{10111010}$ & $\ket{10011000}$ & $\ket{10011000}$ & $\ket{10011000}$ & $\ket{10011000}$ & $\ket{10011000}$ & $\ket{10011000}$
\\

\hline

$\ket{10011010}$ & $\ket{10111020}$ & $\ket{10111120}$ & $\ket{10011110}$ & $\ket{10011110}$ & $\ket{10011110}$ & $\ket{10011110}$ & $\ket{10011110}$ & $\ket{10011010}$
\\
\hline

$\ket{10011100}$ & $\ket{10111110}$ & $\ket{10111110}$ & $\ket{10011100}$ & $\ket{10011100}$ & $\ket{10011100}$ & $\ket{10011100}$ & $\ket{10011100}$ & $\ket{10011100}$
\\
\hline

$\ket{10011110}$ & $\ket{10111120}$ & $\ket{10111220}$ & $\ket{10011210}$ & $\ket{10011210}$ & $\ket{10011210}$ & $\ket{10011210}$ & $\ket{10011210}$ & $\ket{10011110}$
\\

\hline

$\ket{10100000}$ & $\ket{10200000}$ & $\ket{11200000}$ & $\ket{11100000}$ & $\ket{11100000}$ & $\ket{11100000}$ & $\ket{11100000}$ & $\ket{11100000}$ & $\ket{10100000}$
\\
\hline

$\ket{10100010}$ & $\ket{10200010}$ & $\ket{11200010}$ & $\ket{11100010}$ & $\ket{11100010}$ & $\ket{11100010}$ & $\ket{11100010}$ & $\ket{11100010}$ & $\ket{10100010}$
\\

\hline

$\ket{10100100}$ & $\ket{10200100}$ & $\ket{11200100}$ & $\ket{11100100}$ & $\ket{11100100}$ & $\ket{11100100}$ & $\ket{11100100}$ & $\ket{11100100}$ & $\ket{10100100}$
\\

\hline

$\ket{10100110}$ & $\ket{10200110}$ & $\ket{11200110}$ & $\ket{11100110}$ & $\ket{11100110}$ & $\ket{11100110}$ & $\ket{11100110}$ & $\ket{11100110}$ & $\ket{10100110}$
\\
\hline

$\ket{10101000}$ & $\ket{10201010}$ & $\ket{11201010}$ & $\ket{11101000}$ & $\ket{11101000}$ & $\ket{11101000}$ & $\ket{11101000}$ & $\ket{11101000}$ & $\ket{10101000}$
\\

\hline

$\ket{10101010}$ & $\ket{10201020}$ & $\ket{11201120}$ & $\ket{11101110}$ & $\ket{11101110}$ & $\ket{11101110}$ & $\ket{11101110}$ & $\ket{11101110}$ & $\ket{10101010}$
\\

\hline

$\ket{10101100}$ & $\ket{10201110}$ & $\ket{11201110}$ & $\ket{11101100}$ & $\ket{11101100}$ & $\ket{11101100}$ & $\ket{11101100}$ & $\ket{11101100}$ & $\ket{10101100}$
\\

\hline
    \end{tabular}}
    \label{simu}
\end{table*}

\begin{table*}[!htb]
\resizebox{\textwidth}{!}{%
    \centering
    \begin{tabular}{|c|c|c|c|c|c|c|c|c|}
    \hline
    Input State & After $1^{st}$ time cycle & After $2^{nd}$ time cycle & After $3^{rd}$ time cycle & After $4^{th}$ time cycle & After $5^{th}$ time cycle & After $6^{th}$ time cycle & After $7^{th}$ time cycle & Output State after Mirror 
\\
\hline

$\ket{10101110}$ & $\ket{10201120}$ & $\ket{11201220}$ & $\ket{11101210}$ & $\ket{11101210}$ & $\ket{11101210}$ & $\ket{11101210}$ & $\ket{11101210}$ & $\ket{10101110}$
\\

\hline

$\ket{10110000}$ & $\ket{10210000}$ & $\ket{11210000}$ & $\ket{11110000}$ & $\ket{11110000}$ & $\ket{11110000}$ & $\ket{11110000}$ & $\ket{11110000}$ & $\ket{10110000}$
\\

\hline
$\ket{10110010}$ & $\ket{10210010}$ & $\ket{11210010}$ & $\ket{11110010}$ & $\ket{11110010}$ & $\ket{11110010}$ & $\ket{11110010}$ & $\ket{11110010}$ & $\ket{10110010}$
\\
\hline

$\ket{10110100}$ & $\ket{10210100}$ & $\ket{11210100}$ & $\ket{11110100}$ & $\ket{11110100}$ & $\ket{11110100}$ & $\ket{11110100}$ & $\ket{11110100}$ & $\ket{10110100}$
\\
\hline

$\ket{10110110}$ & 
$\ket{10210110}$ & 
$\ket{11210110}$ & 
$\ket{11110110}$ & 
$\ket{11110110}$ & 
$\ket{11110110}$ & 
$\ket{11110110}$ & 
$\ket{11110110}$ & 
$\ket{10110110}$
\\
\hline

$\ket{10111000}$ & $\ket{10211010}$ & $\ket{10211010}$ & $\ket{11211000}$ & $\ket{11111000}$ & $\ket{11111000}$ & $\ket{11111000}$ & $\ket{11111000}$ & $\ket{10111000}$
\\
\hline

$\ket{10111010}$ & $\ket{10211020}$ & $\ket{11211120}$ & $\ket{11111110}$ & $\ket{11111110}$ & $\ket{11111110}$ & $\ket{11111110}$ & $\ket{11111110}$ & $\ket{10111010}$
\\
\hline

$\ket{10111100}$ & $\ket{10211110}$ & $\ket{11211110}$ & $\ket{11111100}$ & $\ket{11111100}$ & $\ket{11111100}$ & $\ket{11111100}$ & $\ket{11111100}$ & $\ket{10111100}$
\\
\hline

$\ket{10111110}$ & $\ket{10211120}$ & $\ket{11211220}$ & $\ket{11111210}$ & $\ket{11111210}$ &
$\ket{11111210}$ &
$\ket{11111210}$ &
$\ket{11111210}$ &
$\ket{10111110}$
\\

\hline

$\ket{11000000}$ & $\ket{11100000}$ & $\ket{11100000}$ & $\ket{11000000}$ & $\ket{11000000}$ & $\ket{11000000}$ & $\ket{11000000}$ & $\ket{11000000}$ & $\ket{11000000}$
\\
\hline

$\ket{11000010}$ & $\ket{11100010}$ & $\ket{11100010}$ & $\ket{11000010}$ & $\ket{11000010}$ & $\ket{11000010}$ & $\ket{11000010}$ & $\ket{11000010}$ &  $\ket{11000010}$
\\
\hline

$\ket{11000100}$ & $\ket{11100100}$ &
$\ket{11100100}$ &
$\ket{11000100}$ & $\ket{11000100}$ & $\ket{11000100}$ & $\ket{11000100}$ & $\ket{11000100}$ & $\ket{11000100}$
\\
\hline

$\ket{11000110}$ & $\ket{11100110}$ & $\ket{11100110}$ & $\ket{11000110}$ & $\ket{11000110}$ & $\ket{11000110}$ & $\ket{11000110}$ & $\ket{11000110}$ & $\ket{11000110}$
\\
\hline

$\ket{11001000}$ & $\ket{11101010}$ & $\ket{11101010}$ & $\ket{11001000}$ & $\ket{11001000}$ & $\ket{11001000}$ & $\ket{11001000}$ & $\ket{11001000}$ & $\ket{11001000}$
\\
\hline

$\ket{11001010}$ & $\ket{11101020}$ & $\ket{11101120}$ & $\ket{11001110}$ & $\ket{11001110}$ & $\ket{11001110}$ & $\ket{11001110}$ & $\ket{11001110}$ & $\ket{11001010}$
\\
\hline

$\ket{11001100}$ & $\ket{11101110}$ & $\ket{11101110}$ & $\ket{11001100}$ & $\ket{11001100}$ & $\ket{11001100}$ & $\ket{11001100}$ & $\ket{11001100}$ & $\ket{11001100}$
\\
\hline

$\ket{11001110}$ & $\ket{11101120}$ & $\ket{11101220}$ & $\ket{11001210}$ & $\ket{11001210}$ & $\ket{11001210}$ & $\ket{11001210}$ & $\ket{11001210}$ & $\ket{11001110}$
\\
\hline

$\ket{11010000}$ & $\ket{11110000}$ & $\ket{11110000}$ & $\ket{11010000}$ & $\ket{11010000}$ & $\ket{11010000}$ & $\ket{11010000}$ & $\ket{11010000}$ & $\ket{11010000}$
\\
\hline

$\ket{11010010}$ & $\ket{11110010}$ & $\ket{11110010}$ & $\ket{11010010}$ & $\ket{11010010}$ & $\ket{11010010}$ & $\ket{11010010}$ & $\ket{11010010}$ & $\ket{11010010}$
\\
\hline

$\ket{11010100}$ & $\ket{11110100}$ & $\ket{11110100}$ & $\ket{11010100}$ & $\ket{11010100}$ & $\ket{11010100}$ & $\ket{11010100}$ & $\ket{11010100}$ & $\ket{11010100}$
\\
\hline

$\ket{11010110}$ & $\ket{11110110}$ & $\ket{11110110}$ & $\ket{11010110}$ & $\ket{11010110}$ & $\ket{11010110}$ & $\ket{11010110}$ & $\ket{11010110}$ & $\ket{11010110}$
\\
\hline

$\ket{11011000}$ & $\ket{11111010}$ & $\ket{11111010}$ & $\ket{11011000}$ & $\ket{11011000}$ & $\ket{11011000}$ & $\ket{11011000}$ & $\ket{11011000}$ & $\ket{11011000}$
\\
\hline

$\ket{11011010}$ & $\ket{11111020}$ & $\ket{11111120}$ & $\ket{11011110}$ & $\ket{11011110}$ & $\ket{11011110}$ & $\ket{11011110}$ & $\ket{11011110}$ & $\ket{11011010}$
\\
\hline

$\ket{11011100}$ & $\ket{11111110}$ & $\ket{11111110}$ & $\ket{11011100}$ & $\ket{11011100}$ & $\ket{11011100}$ & $\ket{11011100}$ & $\ket{11011100}$ & $\ket{11011100}$
\\
\hline

$\ket{11011110}$ & $\ket{11111120}$ & $\ket{11111220}$ & $\ket{11011210}$ & $\ket{11011210}$ & $\ket{11011210}$ & $\ket{11011210}$ & $\ket{11011210}$ & $\ket{11011110}$
\\

\hline

$\ket{11100000}$ & $\ket{11200000}$ & $\ket{12200000}$ & $\ket{12100000}$ & $\ket{12100100}$ & $\ket{12100100}$ & $\ket{12100000}$ & $\ket{12100000}$ & $\ket{11100000}$
\\
\hline

$\ket{11100010}$ & $\ket{11200010}$ & $\ket{12200010}$ & $\ket{12100010}$ & $\ket{12100110}$ & $\ket{12100110}$ & $\ket{12100010}$ & $\ket{12100010}$ & $\ket{11100010}$
\\
\hline

$\ket{11100100}$ & $\ket{11200100}$ & $\ket{12200100}$ & $\ket{12100100}$ & $\ket{12100200}$ & $\ket{12100200}$ & $\ket{12100100}$ & $\ket{12100100}$ & $\ket{11100100}$
\\
\hline

$\ket{11100110}$ & $\ket{11200110}$ & $\ket{12200110}$ & $\ket{12100110}$ & $\ket{12100210}$ & $\ket{12100210}$ & $\ket{12100110}$ & $\ket{12100110}$ & $\ket{11100110}$
\\
\hline

$\ket{11101000}$ & $\ket{11201010}$ & $\ket{12201010}$ & $\ket{12101000}$ & $\ket{12101100}$ & $\ket{12101100}$ & $\ket{12101000}$ & $\ket{12101000}$ & $\ket{11101000}$
\\
\hline

$\ket{11101010}$ & $\ket{11201020}$ & $\ket{12201120}$ & $\ket{12101110}$ & $\ket{12101210}$ & $\ket{12101210}$ & $\ket{12101110}$ & $\ket{12101110}$ & $\ket{11101010}$
\\
\hline

$\ket{11101100}$ & $\ket{11201110}$ & $\ket{12201110}$ & $\ket{12101100}$ & $\ket{12101210}$ & $\ket{12101210}$ & $\ket{12101110}$ & $\ket{12101110}$ & $\ket{11101100}$
\\
\hline

$\ket{11101110}$ & $\ket{11201120}$ & $\ket{12201220}$ & $\ket{12101210}$ & $\ket{12101310}$ & $\ket{12111310}$ & $\ket{12111210}$ & $\ket{12111210}$ & $\ket{11101110}$
\\
\hline

$\ket{11110000}$ & $\ket{11210000}$ & $\ket{12210000}$ & $\ket{12110000}$ & $\ket{12110100}$ & $\ket{12110100}$ & $\ket{12110000}$ & $\ket{12110000}$ & $\ket{11110000}$
\\
\hline

$\ket{11110010}$ & $\ket{11210010}$ & $\ket{12210010}$ & $\ket{12110010}$ & $\ket{12110110}$ & $\ket{12110110}$ & $\ket{12110010}$ & $\ket{12110010}$ & $\ket{11110010}$
\\
\hline

$\ket{11110100}$ & $\ket{11210100}$ & $\ket{12210100}$ & $\ket{12110100}$ & $\ket{12110200}$ & $\ket{12110200}$ & $\ket{12110100}$ & $\ket{12110100}$ & $\ket{11110100}$
\\
\hline

$\ket{11110110}$ & $\ket{11210110}$ & $\ket{12210110}$ & $\ket{12110110}$ & $\ket{12110210}$ & $\ket{12110210}$ & $\ket{12110110}$ & $\ket{12110110}$ & $\ket{11110110}$
\\
\hline

$\ket{11111000}$ & $\ket{11211010}$ & $\ket{12211010}$ & $\ket{12111000}$ & $\ket{12111100}$ & $\ket{12111100}$ & $\ket{12111000}$ & $\ket{12111000}$ & $\ket{11111000}$
\\
\hline

$\ket{11111010}$ & $\ket{11211020}$ & $\ket{12211120}$ & $\ket{12111110}$ & $\ket{12111210}$ & $\ket{12111210}$ & $\ket{12111110}$ & $\ket{12111110}$ & $\ket{11111010}$
\\
\hline

$\ket{11111100}$ & $\ket{11211110}$ & $\ket{12211110}$ & $\ket{12111100}$ & $\ket{12111200}$ & $\ket{12111200}$ & $\ket{12111100}$ & $\ket{12111100}$ & $\ket{11111100}$
\\
\hline

$\pmb{\ket{11111110}}$ & $\pmb{\ket{11211120}}$ & $\pmb{\ket{12211220}}$ & $\pmb{\ket{12111210}}$ & $\pmb{\ket{12111310}}$ & $\pmb{\ket{12121310}}$ & $\pmb{\ket{12121210}}$ & $\pmb{\ket{12121211}}$ & $\pmb{\ket{11111111}}$
\\
\hline
    \end{tabular}}
\end{table*}


\end{document}